\documentclass[journal]{IEEEtran}

\ifCLASSINFOpdf
\else
\fi

\usepackage{subfigure}

\usepackage[ruled]{algorithm2e} 
\usepackage{algorithmicx}
\usepackage{multirow} 
\usepackage{multicol}

\usepackage{lineno}
\usepackage{color,soul}
\usepackage{bm}
\usepackage{amsthm,amsmath,amssymb}
\usepackage{mathrsfs}
\usepackage{overpic}
\usepackage{multirow}
\usepackage{booktabs}       
\usepackage{makecell}
\usepackage{array}
\usepackage{paralist}
\usepackage{subfigure}
\usepackage{graphicx}
\usepackage{pifont}
\usepackage{arydshln}
\modulolinenumbers[5]
\newcommand{\PreserveBackslash}[1]{\let\temp=\\#1\let\\=\temp}
\newcolumntype{C}[1]{>{\PreserveBackslash\centering}p{#1}}
\newcolumntype{R}[1]{>{\PreserveBackslash\raggedleft}p{#1}}
\newcolumntype{L}[1]{>{\PreserveBackslash\raggedright}p{#1}}
\usepackage{stfloats}

\def\revs1#1{{\color{blue}{\it{#1}}}}
\usepackage[colorlinks,            
linkcolor=black,           
anchorcolor=black,            
citecolor=black]{hyperref}
\usepackage{xcolor}
\usepackage{easyReview}
\setreviewsoff

\begin{document}
\title { MorphText: Deep Morphology Regularized Accurate Arbitrary-shape Scene Text Detection}

\author{Chengpei Xu,
        Wenjing Jia,
        Ruomei Wang,
        Xiaonan Luo,
        and Xiangjian He
\thanks{C. Xu, W. Jia, and X. He are with Faculty of Engineering and IT, University of Technology Sydney, Australia (Email: Chengpei.Xu@student.uts.edu.au, Wenjing.Jia@uts.edu.au, Xiangjian.He@uts.edu.au). X. He will be with the Department of Computer Science, University of Nottingham Ningbo China from May 2022 (Xiangjian.He@gmail.com). Corresponding author:  X. He} 
\thanks{R. Wang is with School of Data and Computer Science, National Engineering Research Center of Digital Life, Sun Yat-sen University, China (Email: isswrm@mail.sysu.edu.cn).}
\thanks{X. Luo is with the School of Computer Science and Information Security, Guilin University of Electronic Technology, China (Email: luoxn@guet.edu.cn).}

}

\maketitle

\begin{abstract}
Bottom-up text detection methods play an important role in arbitrary-shape scene text detection but there are two restrictions preventing them from achieving their great potential, \textit{i.e.}, 1) the accumulation of false text segment detections, which affects subsequent processing, and 2) the difficulty of building reliable connections between text segments. 
Targeting these two problems, we propose a novel approach, named ``MorphText", to capture the regularity of texts by embedding deep morphology for arbitrary-shape text detection. 
Towards this end, two deep morphological modules are designed to regularize text segments and determine the linkage between them. 
First, a Deep Morphological Opening (DMOP) module is constructed to remove false text segment detections generated in the feature extraction process. 
Then, a Deep Morphological Closing (DMCL) module is proposed to allow text instances of various shapes to stretch their morphology along their most significant orientation while deriving their connections. 
Extensive experiments conducted on four challenging benchmark datasets (CTW1500, Total-Text, MSRA-TD500 and ICDAR2017) demonstrate that our proposed MorphText outperforms both top-down and bottom-up state-of-the-art arbitrary-shape scene text detection approaches. 
\end{abstract}


\IEEEpeerreviewmaketitle

\section{Introduction}

\maketitle

\IEEEPARstart{A}RBITRARY-shape scene text detection has gained widespread attention in the age of deep learning.
The mainstream approaches for arbitrary-shape scene text detection can be grouped into two types: top-down and bottom-up. 
The top-down approaches usually model the text areas globally and produce the final text areas in a single pass, while bottom-up methods usually model text areas as a connected-component linkage problem. 
The top-down methods align with the current design 
strategy of end-to-end deep learning models, so they often require fewer post-processing steps compared with the bottom-up methods. 
Moreover, some successful modules designed for object detection and semantic/instance segmentation can be directly embedded into top-down methods to enhance detection performance. 
However, texts differ significantly from general objects in terms of scale, aspect ratio, etc. Some direct embedding operations have not considered the characteristics of the texts themselves. 
As a consequence, top-down methods sometimes fail to detect long texts due to the limited reception field of CNNs~\cite{7,16}. 
This has triggered some reception-field expansion methods, such as the non-local~\cite{35} and deformable convolution~\cite{37} modules in~\cite{7,39,43}. 
However, their improvement to arbitrary-shape text detection is limited. 

Bottom-up methods are naturally advantageous for processing long or arbitrary-shape texts, since the rules for forming words and sentences in the real world intrinsically follow a bottom-up manner. 
Bottom-up methods can be more robust in terms of dealing with long texts when the connections between text segments are handled properly. 
Nevertheless, there are two issues that prevent them from achieving their great potential.

\begin{figure}[t]  
\subfigure[GCN-based method~\cite{11}]{
\begin{minipage}[t]{0.46\linewidth}
  
\includegraphics[width=4cm,height=2.4cm]{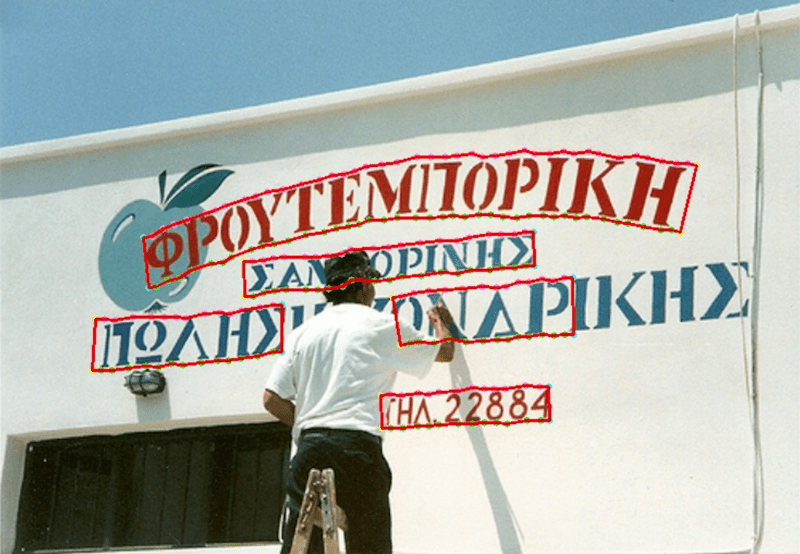}
\label{1a}
\end{minipage}
}               
\subfigure[Top-down method~\cite{9} ]{  
\begin{minipage}[t]{0.46\linewidth}

\includegraphics[width=4cm,height=2.4cm]{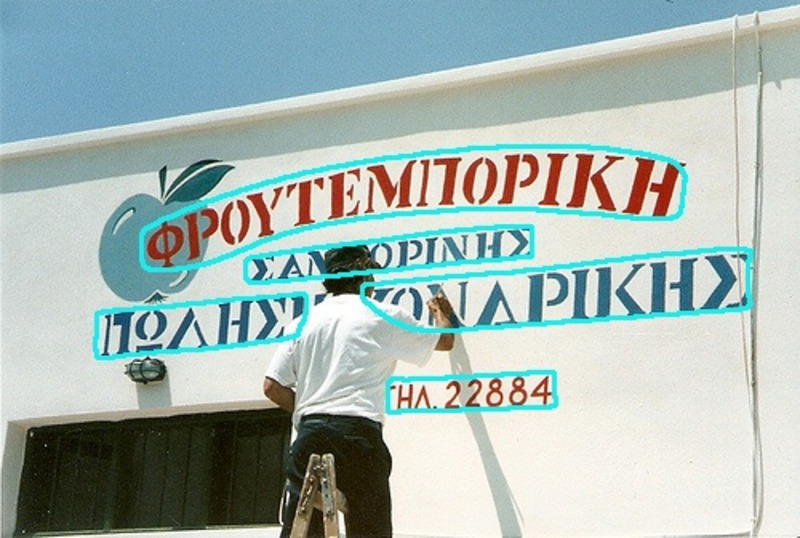}
\label{1b}
\end{minipage}   
} 

\subfigure[Our MorphText]{
\begin{minipage}[t]{0.46\linewidth}
  
\includegraphics[width=4cm,height=2.4cm]{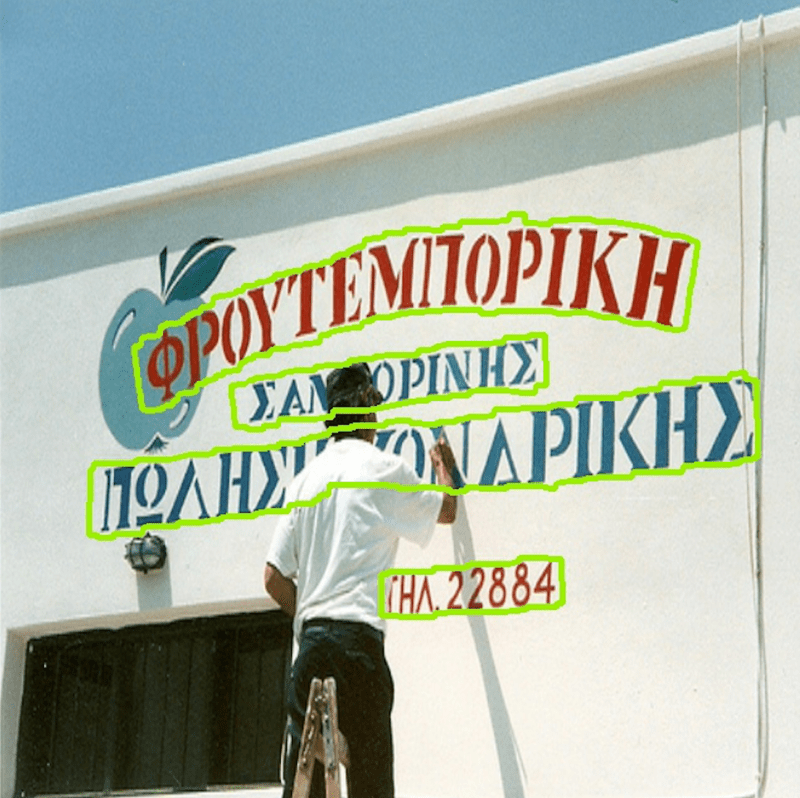}
\label{1c}
\end{minipage}
}               
\subfigure[Ground truth]{  
\begin{minipage}[t]{0.46\linewidth}

\includegraphics[width=4cm,height=2.4cm]{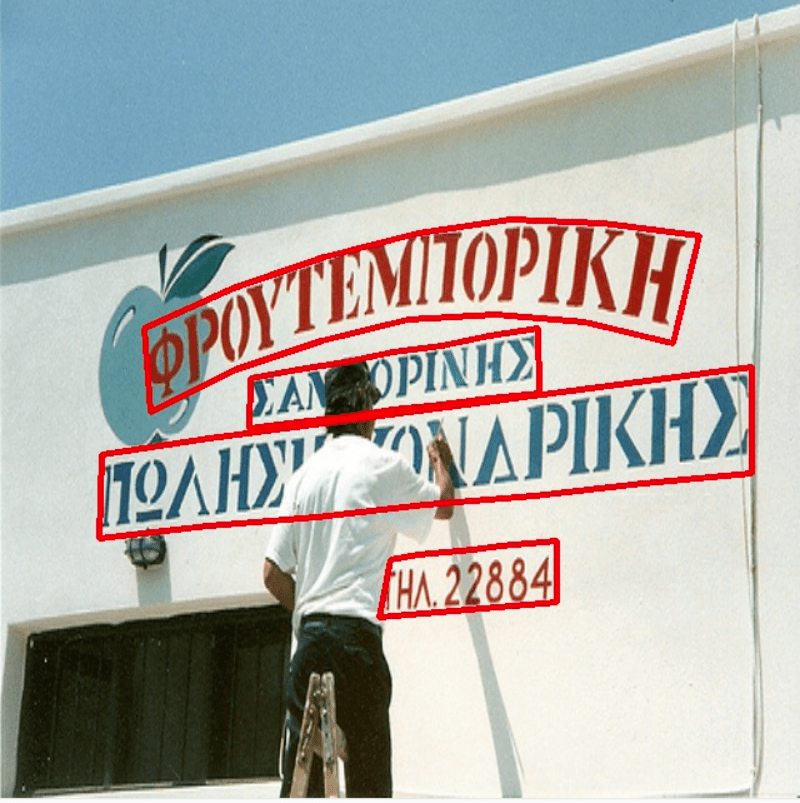}
\label{1d}
\end{minipage}   
} 

\caption{
Both the GCN-based method~\cite{11} and the top-down method~\cite{9} have failed (as shown in (a) and (b)) when, 
the text instance is separated due to heavy occlusion. 
With our morphology regularization, such separated text segments can still be connected into a single text instance (as shown in (c)).} 
\label{2}
\centering
\end{figure}

\begin{figure*}[h]  
\subfigure[The false text segment detection (pink boxes) and missing connections (green boxes) issues of the existing bottom-up approaches]{
\begin{minipage}[t]{1\linewidth}
  
\includegraphics[width=17.5cm, height=3.22cm]{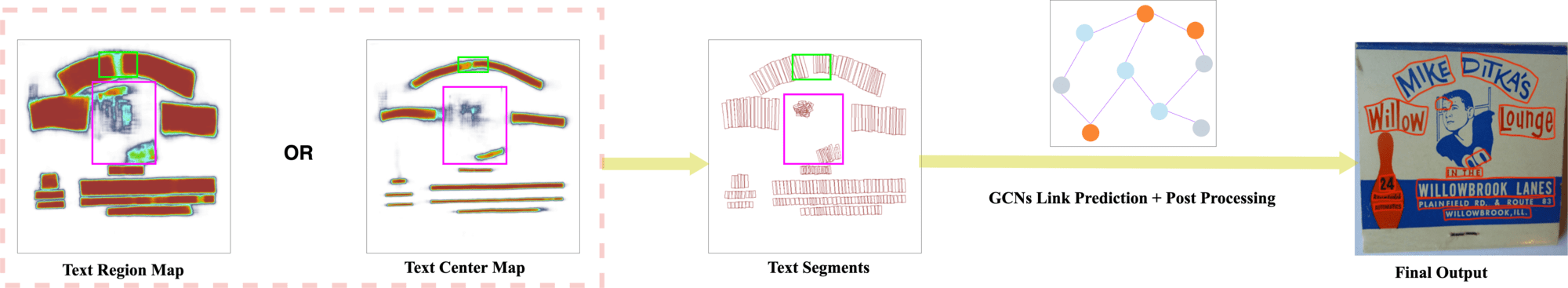}
\label{solve_a}
\end{minipage}
}               
\subfigure[Morphtext ]{  
\begin{minipage}[t]{1\linewidth}

\includegraphics[height=3.5cm]{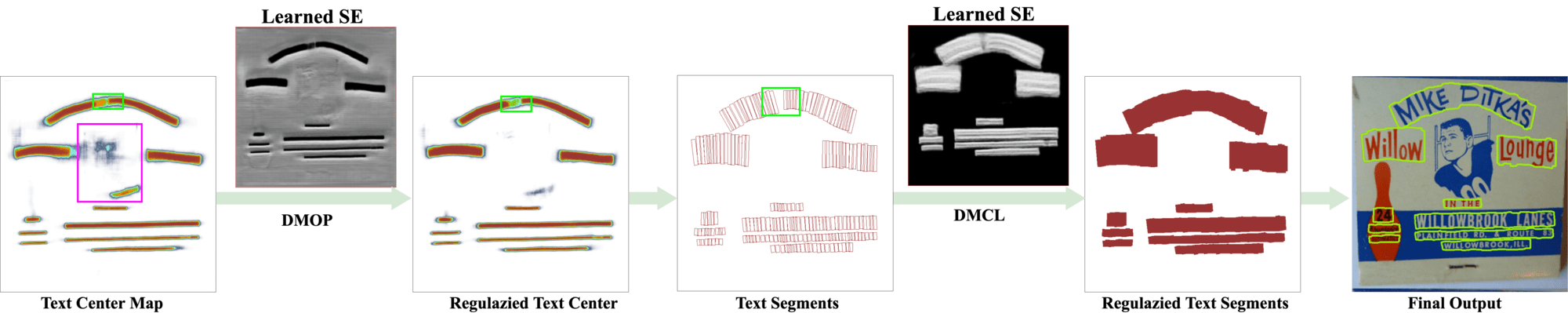}
\label{solve_b}
\end{minipage}   
} 

\caption{Our proposed MorphText approach effectively addresses two key issues that restrain the performance of the bottom-up methods. The pink boxes indicate the false detection areas accumulated from the earlier processing and the green boxes indicate the disconnected areas.} 
\label{fig:issues_solved}
\centering
\end{figure*}

First, the bottom-up methods are more likely to accumulate false positive detections. \add{This} is because bottom-up methods usually need to combine feature representations generated from different modalities, \textit{e.g.}, the visual representation of individual text segments as well as the relational representation between text segments. The processes of learning the visual representation of individual text segments and learning the relational representation between different text segments are usually independent or partially independent 
of each other and are therefore difficult to be optimized simultaneously. 
For example, some state-of-the-art bottom-up methods~\cite{7,8,11} have explored embedding Graph Convolutional Neural Networks (GCNs) for learning the relational information.
However, the training of the visual representation and GCN modules are partially independent, so their gradient update may also be partially independent from each other. 
Also, the cross-modality training sometimes increases the difficulty of training and can be more complicated than the purely CNN-based top-down methods~\cite{9,10,36,45,cse,xu2022semantic}. 
Moreover, the GCN module only makes inferences on the relationship between text segments and makes no selection or rectification on the text segment candidates, so it does not discriminate whether a text segment is a true positive or false positive detection. Undiscriminated false text segments can lead to error accumulation, which affects the performance of the subsequent processing.

The second constraint of bottom-up methods is the difficulty of reasoning the connections between text segments. 
\add{GCNs seem to be an ideal solution for relational reasoning compared to the previous sequence-based, rule-based, RNN-based or geometric-location-based methods in~\cite{3,6,12,21,xu2024seeing, zhang2021rethinking,xu2021s,25, gu2022strokepeo}, since GCNs are able to capture the complex topological structure between text segments.} However, GCNs also require large computational resources for building the topological structure of all text segments. 
GCNs adopted by the bottom-up methods are expected to address the linkage problem of spatially separated text segments, but the performance of GCN link prediction is dependent on the quality of the visual representation of text segments. 

Before generating text segments, the network first needs to predict the entire candidate text area so that text segments can be generated by slicing or regression. However, the segmentation of text may not strictly align with the spatial interval of the text instance, unless the dataset provides character-level annotation. 
Therefore, there are simple cases where the text segments produced by the visual representation stage are clearly spatially adjacent, as well as difficult cases where text segments are spatially separated. 

Thus, on one hand, spatially adjacent text segments (simple cases) are already connected in the original text map. But GCNs unnecessarily compute their connectivity again, adding unnecessary computational burden.
On the other hand, for spatially separated text segments (difficult cases), it is also challenging for GCNs to determine whether the separation is due to the accumulation of errors from the previous CNN stage or the text instance is indeed separated at this point (see Fig.~\ref{1a}). More linkage failure cases from GCN-based methods~\cite{11} are shown in Fig.~\ref{fig:noise and link}.
This is one of the reasons that the performance of GCN-based bottom-up methods~\cite{7,8,11} for arbitrary shape text detection is lower than those of top-down methods~\cite{9,10,42,43}, since GCNs in these methods only perform link prediction while having no ability to prevent error accumulation. \add{Therefore, with additional mechanisms re-checking false text segments, GCNs can be an ideal framework for addressing the linkage problem.}

Is there a method that can remove false detections while simultaneously addressing the linkage problem of bottom-up approaches?

Before the dominance of the deep learning techniques, morphological operations had experienced great prosperity in the areas of image processing and text detection. 
Morphological operations, including erosion, dilation, opening and closing, are designed to analyze the shape and form of an object in the image, so as to filter target patterns and connect distinct patterns in the image. 
They can be directly applied to gray scale images with a predefined Structure Element (SE). 
Compared to CNNs, the efficiency of filtering a specific pattern by a morphological operation is higher as a non-linear function, because CNNs need to stack several layers to accomplish the same task, whereas morphological operations only need to select a proper SE. However, determining the shape, size and flatness of an SE is a process of trial and error and does not always bring performance gains. 
Handcrafted SEs have significant limitations and cannot adapt to different patterns. This is why traditional morphological operations have gradually lost their attraction for image processing tasks. 

Currently, traditional morphological operations have also evolved into Deep Morphological Networks (DMNs) with learnable SEs to capture different patterns in images~\cite{intomo, mondal2019morphological}. 
These methods have also attempted to prove that DMNs can substitute for CNNs as a universal feature extractor or classifier. 
However, existing experiments have shown that this is not an area in which morphology performs well, and have often led to incremental improvement only.  
In our work, we propose utilizing deep morphology in a supplementary role with CNNs to take advantage of their strength in terms of preserving the regular patterns of targets and tackle the linkage problem that linear filters suffer.

Building on the strength of deep morphological operations in preserving patterns of regularity, in this paper we propose a novel arbitrary-shape scene text detection approach to regularize text segments and alleviate the false detection and missing linkage problems of the existing bottom-up methods. Fig.~\ref{fig:issues_solved} shows such an example of handling the existing problems.

In particular, for removing false detection patterns, we design a Deep Morphological Opening (DMOP) module and embed it into the generation process of text segments and their center-line maps so as to suppress the noisy text-line interfering patterns. 
Our proposed DMOP is different from the opening operations in~\cite{intomo, mondal2019morphological}, as its purpose is to regularize the text segment detection results, rather than being used to extract features.

Our proposed embedding of the DMOP module utilizes the core strength of the traditional morphological opening operation combined with trainable kernels to filter those text-like noisy patterns. The figures in Fig.~\ref{fig:issues_solved} illustrate this. 
Thus, our DMOP is designed to regularize the results of CNNs under the guidance of our carefully designed loss functions. 
Moreover, we propose to adopt the residual connection from~\cite{resnet} to ensure our deep morphological opening retains its partial differentiable properties and does not over-process the detection results. 
Thanks to the trainability of the DMOP module, the error-prone post-processing steps used in existing bottom-up methods are unnecessary.

After applying the DMOP regularization, to link different text segments we design a Deep Morphological Closing (DMCL) module, which operates directly on the morphology of each text segment to fill the gaps between them and build the link. 
Here, even if the partial text segments are missing due to insufficient feature representation (\textit{e.g.}, Figs.~\ref{1a} and~\ref{solve_b}), the DMCL module can still have a chance to determine their morphology. 
The DMCL relaxes the restriction of existing bottom-up methods on text segments having fixed geometric forms, allowing text instances of various shapes to stretch in the most significant direction.

The major contributions of this paper are threefold:

1) Towards accurate detection of arbitrary-shape text, we propose a novel MorphText approach by effectively embedding deep morphology for regularizing text segments.

2) Our method enables the overall network to be trained in an end-to-end manner and replaces the error-prone post-processing steps in bottom-up methods with two trainable deep morphological modules.

3) Our method outperforms the state-of-the-art arbitrary-shape text detection methods on several
competitive benchmark datasets. To the best of our knowledge, this is the first time that deep morphology is introduced into this area.

\section{Related Work}

\subsection{Modeling Methods of Arbitrary-shape Scene Text Detection } 

The top-down methods typically consider arbitrary-shape scene text as an instance segmentation problem. 
Therefore, the classic framework Mask-RCNN~\cite{13} and its variations have been widely used, playing an important role in top-down methods, \textit{e.g},  ContourNet~\cite{10}, Xiao \textit{et al.}~\cite{36}, Text-RPN in Wang \textit{et al.}~\cite{46}, TextFuseNet~\cite{9}. 
There are also some segmentation-based methods~\cite{39,16, textpreceptron}, which can be considered as top-down methods since, similar to Mask-RCNN-based methods, these methods also generate mask-like text segmentation maps as their final results. 
In the methods described in~\cite{39,16,textpreceptron}, the text segmentation map is constructed from a text center line and its offset to the boundary. This step is similar to the direct regression of text boundaries and does not require further post-processing steps. 
Thus, top-down methods generally obtain text detection results from the segmentation maps of the region proposal networks.

The bottom-up methods often need to connect text segments or generate text areas progressively from their centers to the border areas. These methods include TextSnake~\cite{21}, CRAFT~\cite{6}, Scribble Lines~\cite{20}, etc. 
Recently, GCNs have been introduced to ensure the linkage between text segments. 
PuzzleNet~\cite{8}, RelaText~\cite{7} and DRRG~\cite{11} are similar works that have utilized GCNs to link text segments. They share the same idea, \textit{i.e.}, first using a text segment proposal network to generate text segments, and then connecting text segments using GCNs. 
However, their methods suffer from the error accumulation problem: if a text-like object is proposed by the text segment proposal network, the GCNs will indiscriminately link this text-like object with other text segments, resulting in error accumulation. 
Furthermore, the bottom-up methods~\cite{8,7,11,20,21, xu2022arbitrary, xu2019lecture2note} often require additional post-processing to determine the visiting order of the contour points. Compared to the top-down methods, they cannot directly generate text regions from the segmentation maps. Although others such as~\cite{psenet, textfield} generate text areas from the inside out with a dilation process, they also require rule-based post-processing to refine the final text contour.

In our approach, we propose to embed deep morphology into arbitrary-shape text detection, aiming to retain the regularity of texts so as to regularize false text segment detection and also address the problem of missing linkages between them. 
Towards this end, two deep morphological modules are designed, both trainable to adapt to different situations. This approach increases the overall integrity and robustness and avoids the tedious post-processing commonly seen in bottom-up methods.

\begin{figure*}[t]
   \includegraphics[width=\linewidth]{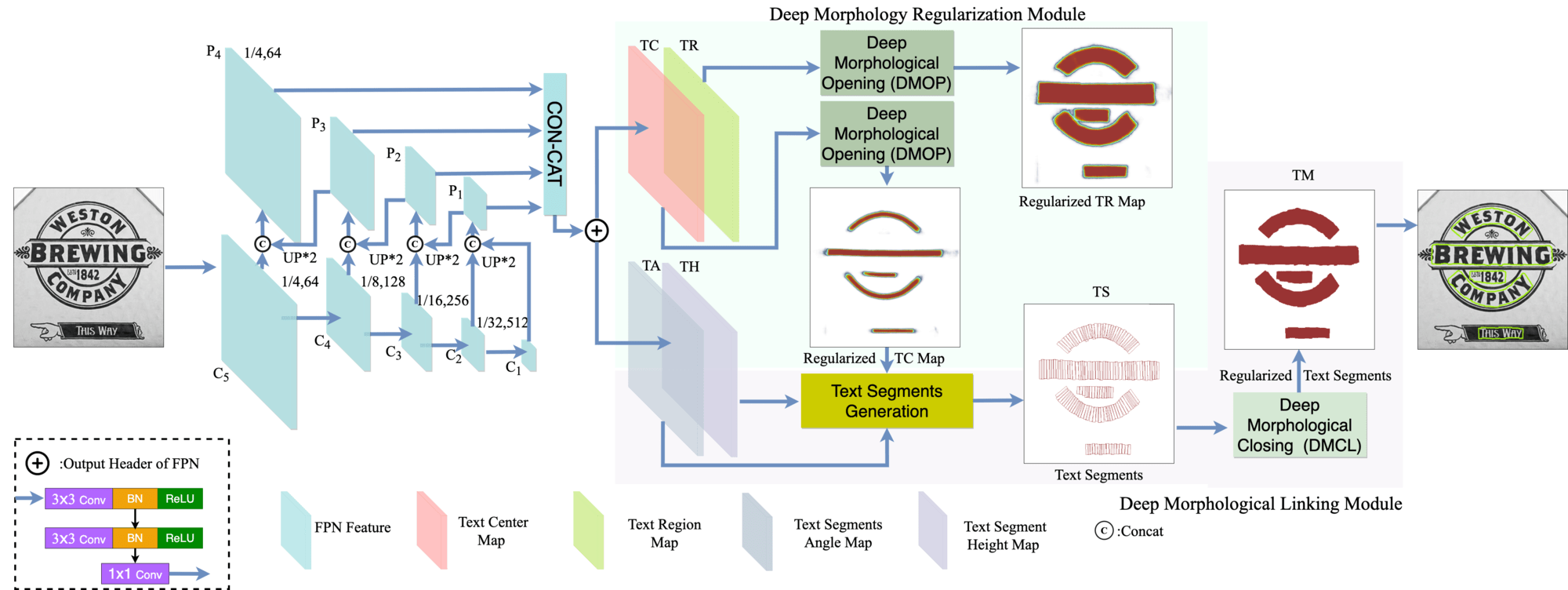} 
  \centering
  \caption{The overall structure of our network, where ``1/4,64", ``1/8,128",... and ``1/32,512" indicate the resize ratio and the channel number. }
  \label{fig:overall_structure}
\end{figure*}
\subsection{Deep Morphological Networks}
The selection of structure elements (SEs) and the enumerative combination of the two basic morphological operations, \textit{i.e.}, erosion and dilation~\cite{mo6} in traditional morphological operations, are the key steps for extracting morphological features and succeeding on the elaborate tasks. Selecting SEs and the combination of morphological operations manually requires expert knowledge and yet the robustness of these handcrafted steps cannot be guaranteed.
Thus, it is logical to consider the idea of learning the SEs automatically and this has become feasible during the current deep learning era~\cite{mo7}. Zamora \textit{et al.}~\cite{mo7} trained morphological neurons with a one dimensional SE using a fully connected neural network. Recently, deep morphological methods~\cite{intomo,derain,prmo,mondal2019morphological} extended the representation of SEs to two dimensions, and enabled 2D images to be used as input for deep morphological networks. Among them, Nogueira \textit{et al.}~\cite{intomo} used a flat SE to build morphological neurons for extracting features as a replacement for linear convolutions. 
However, flat SEs only define the shape of the filtering process and do not affect pixel values, so they may lead to insufficient feature representation on images. 
Later, trainable non-flat SEs were introduced in~\cite{prmo} where a morphology-based pooling method was developed to replace the max pooling process in CNNs. 
Mondal \textit{et al.}~\cite{derain, mondal2019morphological} also adopted trainable non-flat SEs and proved that an alternative stacking of deep morphological erosion and dilation could help with image restoration without needing CNNs.

Although the methods in~\cite{intomo, derain, mondal2019morphological} showed that deep morphological networks contain far fewer parameters than CNNs, it is difficult for deep morphological networks to compete with CNNs in terms of feature extraction and classification capabilities. 
This conclusion can be drawn if we look at the results of image dehazing, image classification and de-raining tasks in~\cite{derain, mondal2019morphological,intomo}.
The min/max algebra in the morphological operation can only be piece-wise differentiable during training, resulting in the inability to precisely and progressively update the gradient at each pixel when compared with the back-propagation process in CNNs. 
Moreover, the min/max algebra is similar to the max pooling layer in CNNs and it inevitably results in the loss of information.

\section{Methodology}

The architecture of our method, illustrated in Fig.~\ref{fig:overall_structure}, contains three parts, namely: a text segment proposal module; a deep morphology-based regularization module; and a morphological linking module. 
Text images are first fed into the text segment proposal module, which uses ResNet50~\cite{resnet} as the backbone and FPN~\cite{5} as the neck to extract multi-scale features of texts. 
Then, the multi-scale features are concatenated and processed by the output head. 
The output head generates a text probability map and its center probability map indicating the center points of text segments, and two probability maps indicating the height and rotation angle of text segments. 
The DMOP module regularizes the noisy text-like patterns in the text and text center maps with learned SEs and removes possible false text segments. 
The regularized text center map is then used to guide the fusion of the text segment height and angle maps and produce candidate text segments. 
Finally, the DMCL module processes the resultant text segments, determining the connections between them based on their morphology and producing the final result. 

In the following sub-sections, we describe the two deep morphology modules and then give details of the text segment proposal module and the objective function.

\subsection{Deep Morphological Operations}

The deep morphological operations in our approach perform similar functions to the traditional image morphological operations but with non-flat, trainable structure elements~\cite{mo6, derain,mo9}. 
Moreover, the input of our deep morphology modules is multiple-channel image features generated from the text segment proposal module and FPN, rather than gray scale images, as indicated in Fig.~\ref{fig:overall_structure}.

Let us denote the multi-channel image feature of size $C \times\ W \times H $ by $\mathcal{I}$, where $C$, $W$ and $H$ are the channel, width and height of the input image features, respectively. 
Similarly, let us denote the size of the structure element $S$ by $C \times M \times N$, where $C$, $M$ and $N$ are the channel, width and height of the structure element, respectively. 

The two basic image morphological operations, \textit{i.e.}, dilation ($\oplus$) and erosion ($\ominus$), can then be expressed by the operation of the structure element $S$ on the input $I$ as:  
\begin{equation}
\label{eq11}
(\mathcal{I}\oplus S)[c,w,h]\!\! =\!\!\max_{(i,j) \in D_s}({
   \mathcal{I}(c,w+i,h+j)+S(c,i,j)}),  
 \end{equation}
 and 
 \begin{equation}
 \label{eq22}
(\mathcal{I}\ominus S)[c,w,h]\!\! = \!\!\min_{(i,j) \in D_s}({
   \mathcal{I}(c,w+i,h+j)-S(c,i,j)}),    
\end{equation}
where, $c \in [1,C]$,  $w \in [1,W]$, $h \in [1,H]$, and 
$D_s$ is the definition field of the structure element $S$, which can be determined by $M$ and $N$. 
In Eqs.~(\ref{eq11}) and~(\ref{eq22}), $\mathcal{I}(c,w+i,h+j)$ means that the structure element $S$ slides on the $c$-th channel of the image feature $\mathcal{I}$, in a similar manner to the convolution operation in CNNs. 
In our experiments, the structure elements for dilation and erosion are initialized as zero matrices, which are then updated 
through optimizing the objective function.

\subsection{Deep Morphology based Text Segment Regularization}

False text segments affect the accuracy of both bottom-up and top-down methods. 
False detection means that text-like interfering patterns are also included in the resultant text segments.

\begin{figure}[t]
  \includegraphics[width=\linewidth]{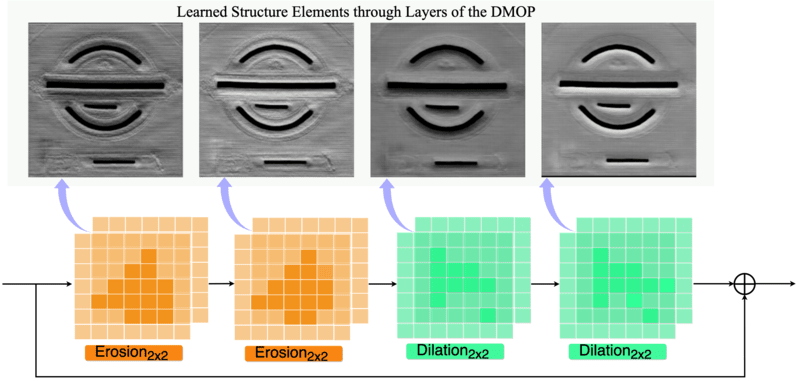}
  \centering
  \caption{The visualization of the learned structure elements through DMOP. } 
  \label{fig:DMOP}
\end{figure}

The traditional morphological opening operation~\cite{mo6} is an erosion operation followed by a dilation operation, and it can be used for removing small, irregular objects from detection. 
The basic principle of our design is also a series of deep erosion operations followed by a series of deep dilation operations. 
Fig.~\ref{fig:DMOP} shows the structure of our Deep Morphological Opening (DMOP) module, which is composed of two erosion blocks followed by two dilation blocks. 
As can be seen in Fig.~\ref{fig:DMOP}, the learned SEs gradually suppress the noise patterns with the deep morphological operations. 
Here, the learned SEs are obtained by splicing the structure elements according to their positions in the image feature.

Considering an input text center map $TC$, the text center map after applying the DMOP module, denoted by $TC^{\prime}$, can be written as:
\begin{equation}
\label {eq:tc:prime}
TC^{\prime}=(TC\ominus S_1^{2\times2\times2} \ominus S_2^{2\times2\times2}) \oplus S_3^{2\times2\times2}\oplus S_4^{2\times2\times2},
\end{equation}
where $S_1^{2 \times 2\times2}$ to $S_4^{2 \times 2\times2}$ are trainable structure elements with size $2\times2$ of the same channels as the text center maps. 
The text region map is processed by DMOP in a similar way.

Note that, a major difference between our DMOP and the deep morphological methods in~\cite{derain, mondal2019morphological,intomo, prmo} is that we introduce the residual connection. Since the min/max operation in erosion and dilation are piece-wise differentiable, the gradient descent of the min/max values will be updated in SEs during the training process. 
For instance, if we consider $L$ as the loss of the structure element $S$ for Eq.~\eqref{eq11}, the gradient update of $S$ can be represented by $\frac{\partial L}{\partial S}$, which can be piece-wise calculated according to the chain rule:
\begin{equation}
\label{eq5}
\frac{\partial L}{\partial S} = \sum_{(i,j) \in D_s} 
\frac{\partial(\mathcal{I}\oplus S [c,w,h])}{\partial S(c,i,j)} 
\cdot 
\frac{\partial L}{\partial (\mathcal{I}\oplus S [c,w,h])}. 
\end{equation}
According to the max/min algebra, 
\begin{equation}
\label{eq6}
\frac{\partial(\mathcal{I}\oplus S [c,w,h])}{\partial S(c,i,j)}=0 \quad or \quad 1.
\end{equation}
That is to say that Eq.~\eqref{eq6} equates one if and only if $\mathcal{I}(c,w+i,h+j)+S(c,i,j)$ reaches the maximum. 
Thus, when the expression in Eq.~\eqref{eq6} is equal to zero it causes the gradient update of the SE to ignore the non-min/max values and not be constrained. Therefore, when the SE removes true positive text areas and causes over-correction, the back-propagation from non-min/max values should also be preserved to compensate for the gradient update.

In our approach, we propose introducing a residual connection into the deep morphology regularization module, where the regularized text center map, denoted by $TC^{\circ}$, after applying DMOP regularization, is computed as:
\begin{equation}
\label {eq:tc:residual}
TC^{\circ }=TC+TC^{\prime},
\end{equation}
where $TC^{\prime}$ is computed by Eq.~(\ref{eq:tc:prime}).

Furthermore, different from the methods in~\cite{derain, mondal2019morphological,intomo, prmo}, which tried to concatenate a sequence of erosion and dilation and their dual sequence and used them as a classifier or feature extractor, our method focuses on using the deep morphological operation as a regularization module. 
Thus, we select small SEs and shallow deep morphological layers to constrain the influence of the DMOP module onto regularizing false detection only. 
Section~\ref{subsect:ablationstudies} shows ablation studies validating the effectiveness of the selection of the SE size and the effect of different sizes and ways of combination.

\subsection{Deep Morphology based Relational Reasoning}
 \add{The center location of text segments $TS$ can now be positioned using $TC^{\circ}$. The other geometric features of text segments are described in Sect.~\ref{ts proposal}. }
After the text segments are obtained, some of the existing bottom-up methods try to firstly determine the linkage relationship between them and then the visiting order of their contour points. 
However, if we look at the text segments in Fig.~\ref{fig:overall_structure}, their shapes and contours are already very close to the ground truth text areas. 
The only problem that we need to solve is if the intervals between text segments should be retained and if the holes in each text segment should be filled up. 

\begin{figure}[t]
  \includegraphics[width=\linewidth]{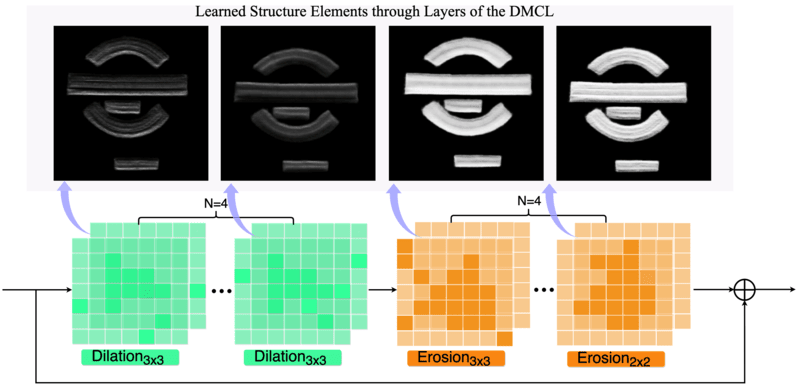}
  \centering
  \caption{Visualization of the learned structure elements through DMCL. }
  \label{fig:DMCL}
\end{figure}

The traditional morphological closing operation~\cite{mo6}, a dilation operation followed by an erosion operation, has been used to fill small holes and connect some patterns in images. 
Similar to the DMOP, we propose a Deep Morphological Closing (DMCL) module, which uses four deep morphological dilation operations followed by four erosion operations to build the link between text segments and fill the holes in them (see Fig.~\ref{fig:DMCL}). 
Similarly, to avoid the over-correction problem, the residual connection is also adopted in our DMCL module.

Thus, the regularized text segment map, denoted by $TS^{\bullet }$, after applying the DMCL module can be represented by:
\begin{equation}
\add{TS^{\bullet }=TS+TS^{\prime \prime}},
\end{equation}
where
\begin{equation}
TS^{\prime\prime}=TS^{\prime}\ominus S_5^{1\times3\times3} \ominus S_6^{1\times3\times3} \ominus S_7^{1\times3\times3} \ominus S_8^{1\times3\times3}. 
\end{equation}
In this equation, 
\begin{equation}
TS^{\prime}=TS\oplus S_1^{1\times3\times3} \oplus S_2^{1\times3\times3} \oplus S_3^{1\times3\times3} \oplus S_4^{1\times3\times3}.
\end{equation}
Here, the size of the SEs is set to $3\times3$ empirically.
The stacking of four deep morphological layers is also determined empirically to fill small holes in the text segment maps. 
The learned SEs in Fig.~\ref{fig:DMCL} show the process of text segments being reshaped and filled to a smooth text instance.

After applying the DMCL module, the text detection results can be directly found in the text segment maps with the hole and connection problem fixed. 
Thus, our DMCL introduces trainable SEs to build the link and fill the holes, and to replace the tedious post-processing steps, such as those in methods~\cite{7,8,11,10}. Our method does not need to locate the contour points or identify their visiting order. The contour drawing process now is the same as in top-down methods. Thus, post-processing steps such as segment grouping and false detection removal  are packed into an end-to-end manner.

\subsection{Text Segment Proposal Module}
\label{ts proposal}
The proposed text segment proposals in our work are rectangles with rotation angles. Each text segment can be represented by $[x,y,h,w,\theta]$, namely, the $x$ and $y$ coordinates of the center points, the height, width and rotation angle of the text segment. 
Here, we concatenate the FPN features $P1-P4$ and use an output head (see Fig.~\ref{fig:overall_structure}) to obtain the 6-channel geometric features $GF$ of the text segments, where $GF \in \mathbb{R}^{6\times W \times H}$. 
The $GF$ contains four text probability maps, namely, two channels for the Text Region (TR) map, two channels for the Text Center (TC) map, one channel for the Text Height (TH) map and one channel for Text Angle (TA) map, \textit{i.e.}, TR$\in\mathbb{R}^{2\times W \times H}$, TC$\in\mathbb{R}^{2\times W \times H}$, TH$\in\mathbb{R}^{1\times W \times H}$ and TA$\in\mathbb{R}^{1\times W \times H}$. 
Among them, the TR map is not directly involved in the calculation of text segments but is used to guide the training of the FPN feature extractor. 
The ground truth of TC maps is generated by shrinking the polygon using the clipping algorithm proposed in~\cite{vatti1992generic}, where the shrinking ratio is set to 0.2. 

To generate the ground truth of TH and TA maps, we first use the method in~\cite{27} to find the bottom long side of the polygon representation of each text region. 
Then, the bottom long side is sectioned to $n$ line segments with $n-1$ points.
Here, for every four pixels we sample a point. 
To generate the TH map, for each point in the TC map, we find the distance from this point to its closest line segment and two times the distance is used to approximate the height of this point. Then, we calculate the angle of the distance line of each point in the TC map with a horizontal line to obtain the TA map. 

Note that $GF$ does not produce the feature map for the width of the text segments. 
This is because our DMOP module has trainable SEs, so that it is not necessary to predict the exact width of the text segments, 
as each text segment will go through the dilatation operation in the DMOP module. 
Thus, in our approach, we simply calculate the Text Width (TW) map based on the resultant TH map. 
Moreover, we design the text segments as slender rectangles, mimicking the general shape of characters. 
Thus, the TW map is calculated by:
\begin{equation}
\label{TW}
TW=Clip(2,\frac{1}{4}TH,8),
\end{equation}
where $Clip$ is the clipping function that limits the values in the $TW$ map to the range of $2 - 8$ pixels. 

Then, with TC, TH, TW and TA, the TS can be generated for the DMOP module.
For every point in the TC map, a rectangle text segment is generated where text segments with over 50\% overlapping are removed by non-maximum suppression to reduce unnecessary computation. 

\subsection{Objective Function}

The overall objective function consists of five parts:
\begin{equation}
L=\lambda_1 L_{TR}+ \lambda_2 L_{TC}+ \lambda_3 L_{TH}+ \lambda_4 L_{TA}+\lambda_5 L_{TM}, 
\end{equation}
where $L_{TR}$ and $L_{TC}$ are the losses from the regularized TR and TC maps, and $L_{TM}$ is the loss from text segments after the morphological operation. 
They are all class-balanced cross-entropy losses for guiding the training of text areas and their center line areas. $L_{TM}$ can be calculated using the same loss function as TR. Online hard example mining~\cite{40} is adopted for training the $L_{TR}$ and $L_{TM}$ functions, keeping the ratio of positive sample number to negative sample number to $1:3$. \add{$L_{TC}$ is also a class-balanced cross-entropy loss with the balancing factor between positive samples and negative samples 
being set to 0.75 empirically to address the unbalanced-sample problem between the text pixels and background pixels.}  $L_{TH}$ and $L_{TA}$ are Smoothed-L1 losses for the height and angle of each text segment. In our paper, $\lambda_1$, $\lambda_2$, $\lambda_3$, $\lambda_4$ and  $\lambda_5$ are set to 1, 2, 1, 1 and 1, respectively, for simplicity.

\begin{figure*}[t]  

\vskip -12pt
\subfigure{
\begin{minipage}[t]{0.18\linewidth}
  
\includegraphics[width=3.4cm,height=3cm]{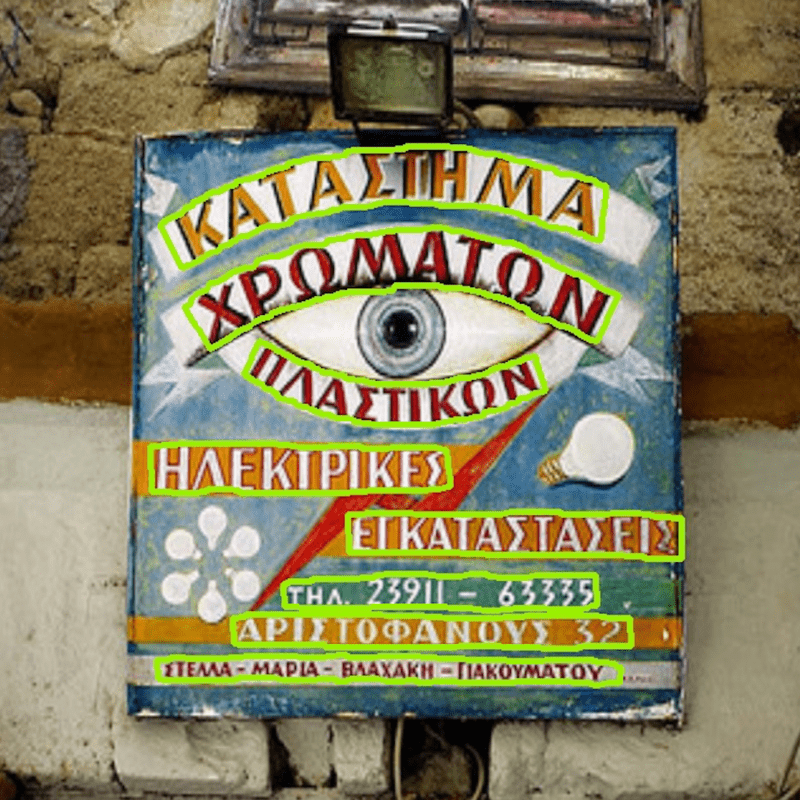}
\label{2a}
\end{minipage}
}               
\subfigure{  
\begin{minipage}[t]{0.18\linewidth}

\includegraphics[width=3.4cm,height=3cm]{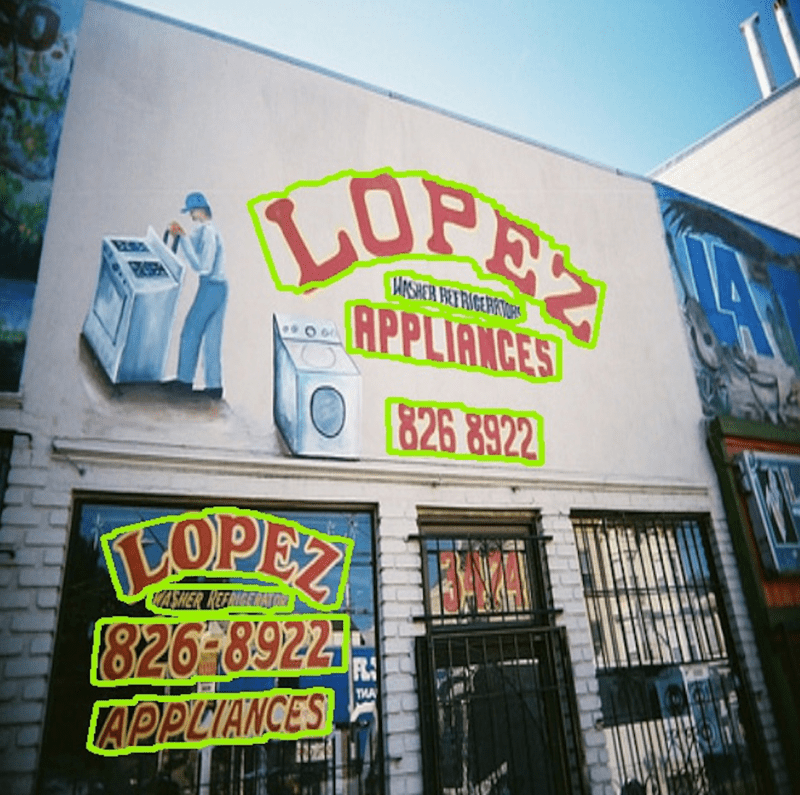}
\label{2b}
\end{minipage}   
} 
\subfigure{  
\begin{minipage}[t]{0.18\linewidth}

\includegraphics[width=3.4cm,height=3cm]{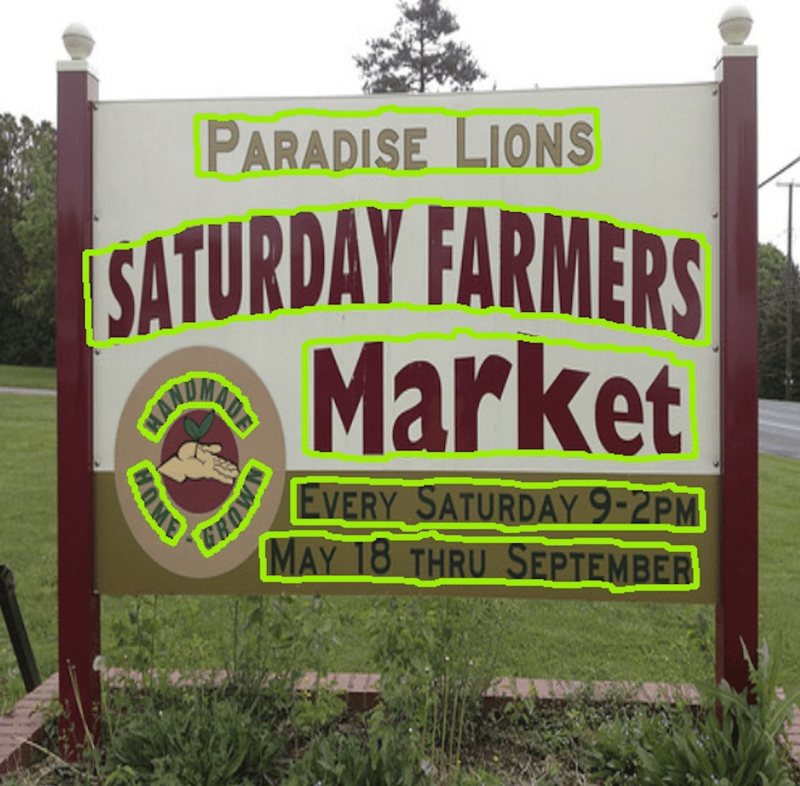}
\label{2b}
\end{minipage}   
} 
\subfigure{  
\begin{minipage}[t]{0.18\linewidth}

\includegraphics[width=3.4cm,height=3cm]{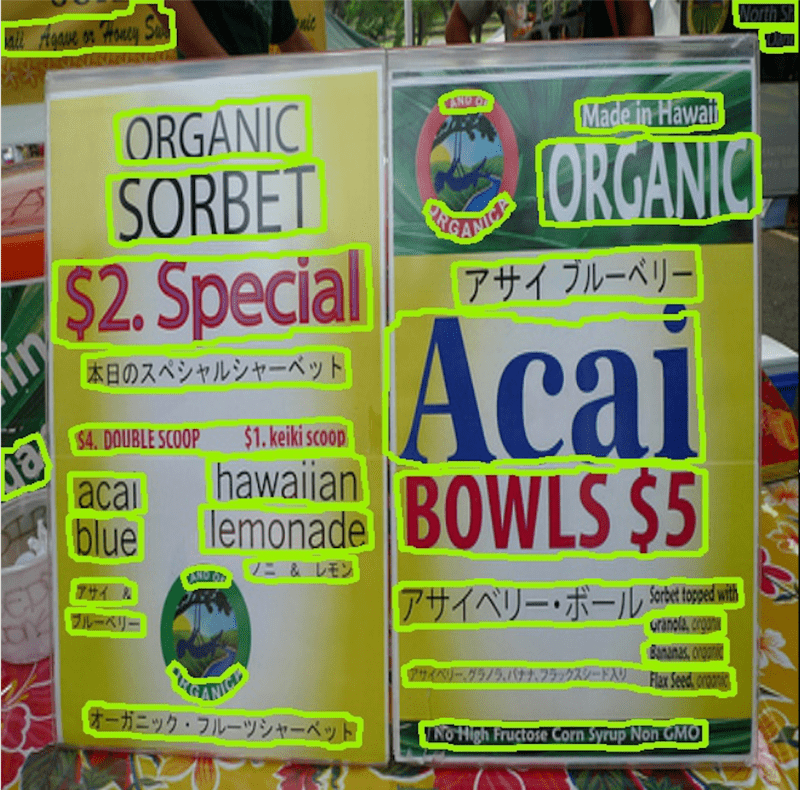}
\label{2b}
\end{minipage}   
} 
\subfigure{  
\begin{minipage}[t]{0.18\linewidth}

\includegraphics[width=3.4cm,height=3cm]{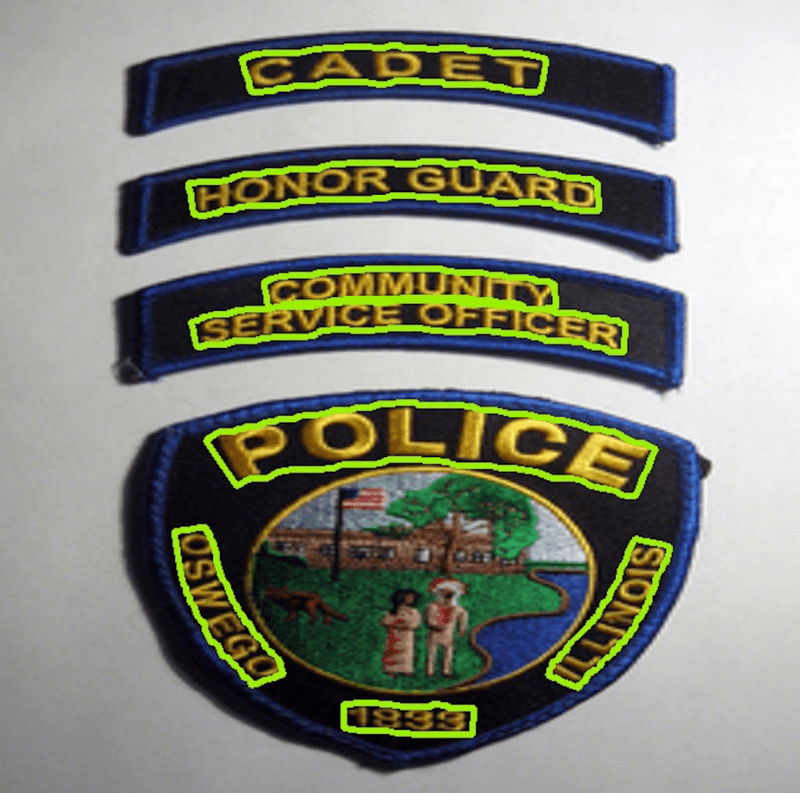}
\label{2b}
\end{minipage}   
} \vskip -12pt
\subfigure{
\begin{minipage}[t]{0.18\linewidth}
  
\includegraphics[width=3.4cm,height=3cm]{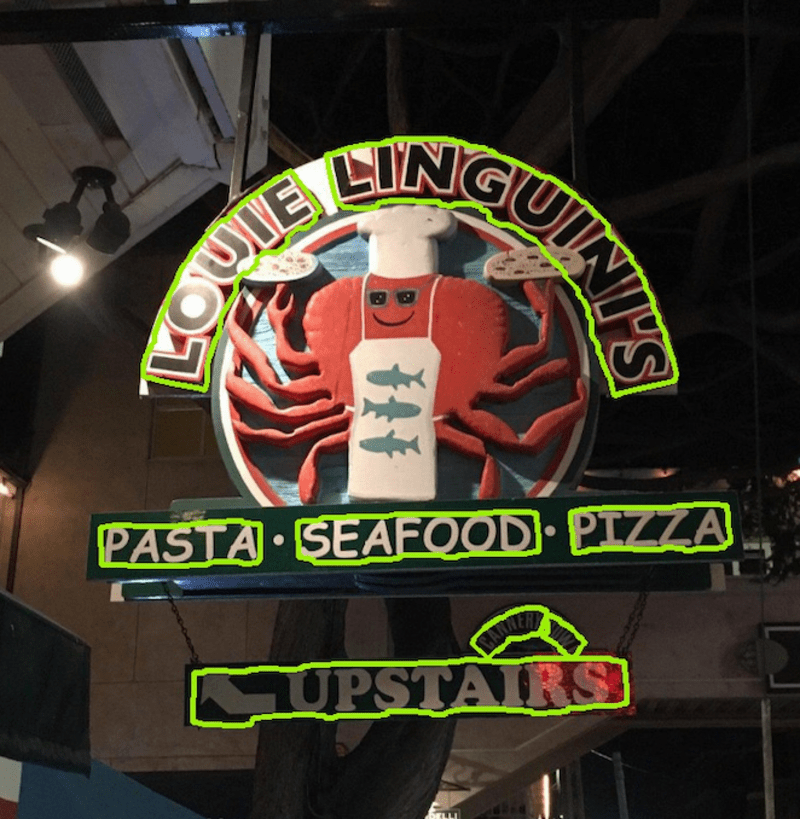}
\label{2a}
\end{minipage}
}               
\subfigure{  
\begin{minipage}[t]{0.18\linewidth}

\includegraphics[width=3.4cm,height=3cm]{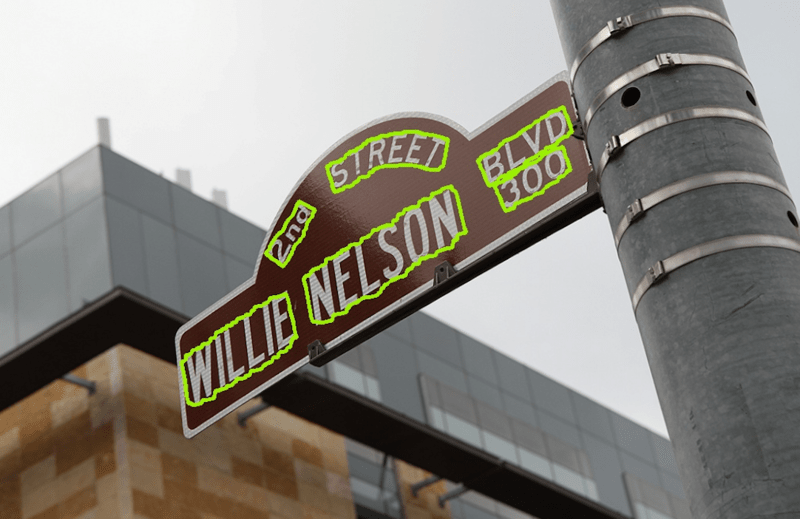}
\label{2b}
\end{minipage}   
} 
\subfigure{  
\begin{minipage}[t]{0.18\linewidth}

\includegraphics[width=3.4cm,height=3cm]{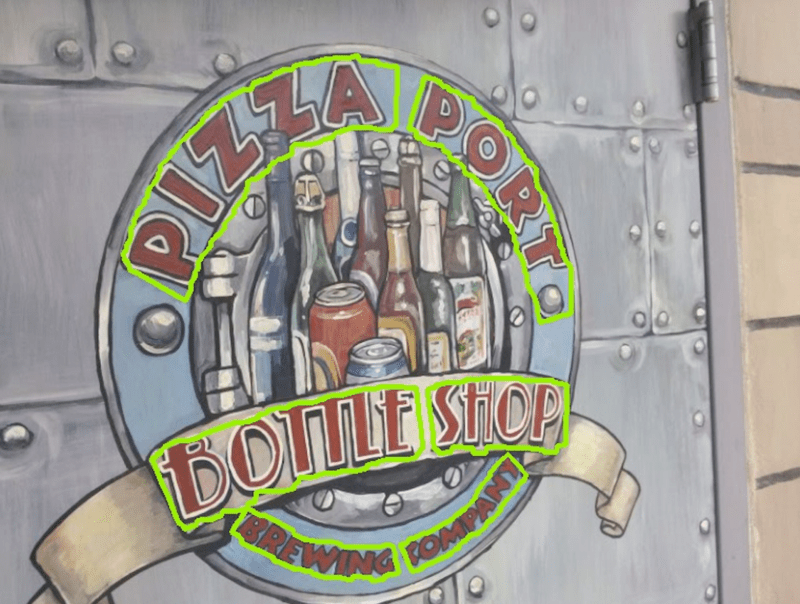}
\label{2b}
\end{minipage}   
} 
\subfigure{  
\begin{minipage}[t]{0.18\linewidth}

\includegraphics[width=3.4cm,height=3cm]{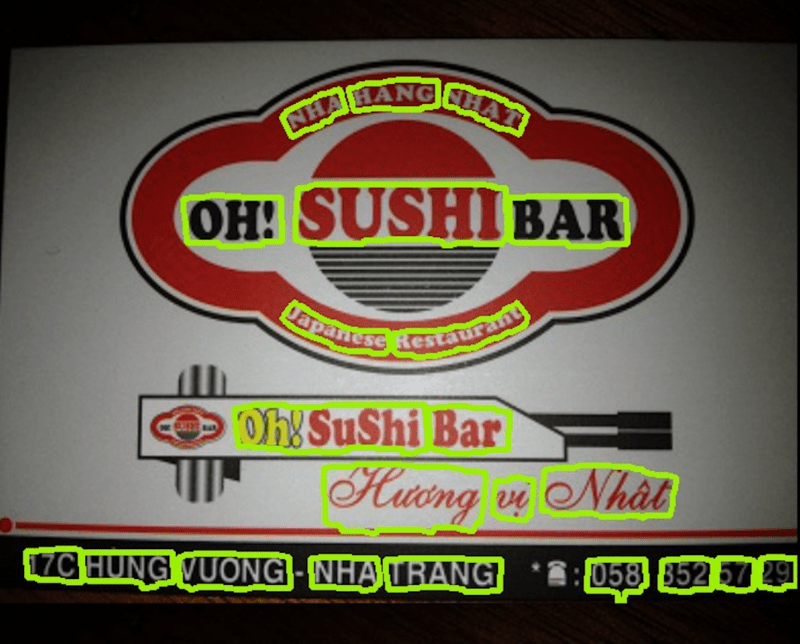}
\label{2b}
\end{minipage}   
} 
\subfigure{  
\begin{minipage}[t]{0.18\linewidth}

\includegraphics[width=3.4cm,height=3cm]{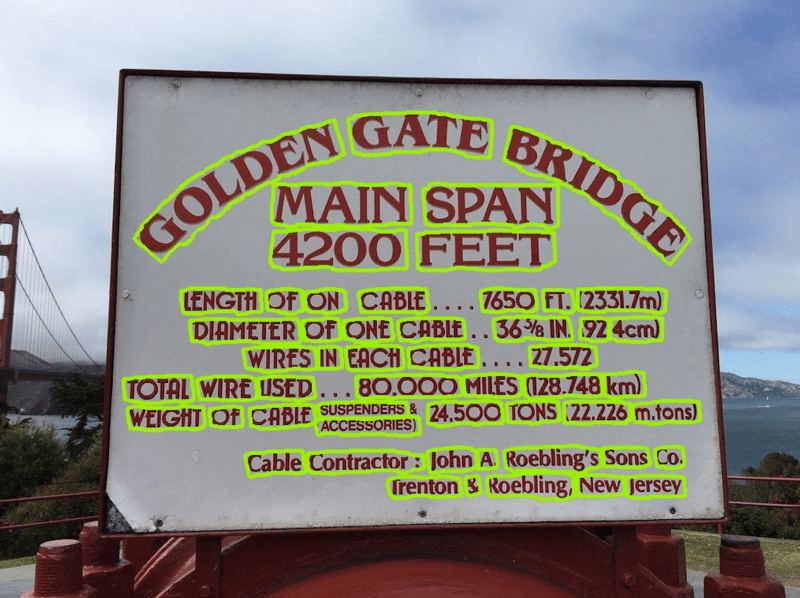}
\label{2b}
\end{minipage}   
} \vskip -12pt
\subfigure{
\begin{minipage}[t]{0.18\linewidth}

\end{minipage}   
} 

\vskip -10pt
\centering
\caption{Examples of text detection results obtained with the proposed MorphText approach on benchmark datasets.} 
\label{fig:visualresults}
\end{figure*}

\section{EXPERIMENTS}
\label {sect:exp}

\subsection{Datasets and Evaluation Protocol}

\noindent{\textbf{\textit{SynthText}}: The SynthText dataset~\cite{synthetic}} consists of 800,000 synthetic images, created by mixing rendered words on natural images. In our work, we follow~\cite{29,6,39,11,7} and use SynthText as a pre-train dataset.\\
 \textbf{\textit{CTW1500}}: The CTW1500 dataset~\cite{19} consists of arbitrary-shape English and Chinese text instances and is a challenging dataset for long curved texts, with 1,500 training and 1,000 testing images. \\ 
 \textbf{\textit{Total-Text:}} The Total-Text dataset~\cite{18}, consisting of 1,255 training images and 300 testing images, was created for detecting arbitrary-shape texts and consists of multi-oriented and curved text instances with polygon annotations. \\
 \textbf{\textit{MSRA-TD500:}} The MSRA-TD500 dataset~\cite{TD500} is designed for detecting multi-oriented and multi-lingual long texts. It contains 300 training images and 200 testing images with line-level annotation. \\
 \textbf{\textit{ICDAR2017:}} The ICDAR2017 dataset~\cite{MLT} is designed for detecting multi-oriented and multi-lingual texts of nine languages with 7,200 training images and 1,800 validation images and 9,000 testing images. Following~\cite{42,psenet,tmm3,11,36}, we use ICDAR2017 as part of our pre-train data.

\begin{table}[h]
  \caption{The effectiveness of our proposed DMOP and DMCL on CTW1500 and Total-Text.}
  \label{tab:1}
   \centering
    \setlength\tabcolsep{5pt}
  \begin{tabular}{c|c|c|c|c|c|c|c}
    \hline
    Datasets& OP& CL & DMOP &DMCL &P (\%) &R (\%) &F (\%)\\
     \hline
    $\quad$&$ \times $&$\checkmark$&$\times$&$\times$&85.7 &82.7& 84.2\\
  
    $\quad$&$\checkmark$&$\checkmark$&$\times$&$\times$&86.5 &82.9& 84.7\\
  
    CTW1500&$\times$&$\checkmark$&$\checkmark$&$\times$&87.2 &83.0& 85.0\\
  
    $\quad$&$\times$&$\times$&$\times$&$\checkmark$&87.6 &{\bfseries83.3}& 85.4\\
    
    $\quad$&$\checkmark$&$\times$&$\times$&$\checkmark$&88.0 &83.2& 85.5\\
  
    $\quad$&$\times$&$\times$&$\checkmark$&$\checkmark$&{\bfseries89.0} &83.2 &{\bfseries86.0}\\
    \hline
    \hline
    $\quad$&$ \times $&$\checkmark$&$\times$&$\times$&86.5 &84.8& 85.6\\
    
    $\quad$&$\checkmark$&$\checkmark$&$\times$&$\times$&87.1 &85.2& 86.1 \\
    
    Total-Text&$\times$&$\checkmark$&$\checkmark$&$\times$&87.9 &84.8& 86.3\\
    
    $\quad$&$\times$&$\times$&$\times$&$\checkmark$&87.4 &85.5&86.4\\
    
    $\quad$&$\checkmark$&$\times$&$\times$&$\checkmark$&87.7 &85.5&86.6\\
        
    $\quad$&$\times$&$\times$&$\checkmark$&$\checkmark$&{\bfseries88.4} &{\bfseries85.5}& {\bfseries86.9}\\
  \hline
\end{tabular}
\end{table} 
\subsection{Implementation Details}
Our MorphText is implemented using the PyTorch 1.7 framework. The backbone network ResNet50 is pre-trained on ImageNet, with FPN being adopted for multi-scale feature extraction. The channels of the up-sampling pipeline are set to 512, 256, 128 and 64 respectively.

The training of MorphText can be divided into two steps. First, we pre-trained our model on SynthText for five epochs using the Adam optimizer with a learning rate of $0.001$. The input image size for pre-training was set to $640\times 640$ pixels. Then, before we fine-tuned our model on specific datasets, ICDAR2017 was used as an additional pre-train dataset for fine-tuning for another 100 epochs. 
We adopted the SGD optimizer and an initial learning rate of $0.01$, which was decayed by a factor 0.1 for every 100 epochs. The momentum and weight decay of SGD were set to $0.9$ and $ 0.0001$, respectively. Following the existing practice, the input size was set to $640\times 640$ for CTW1500 and Total-Text and $960\times 960$ for MSRA-TD500 and ICDAR2017. 
The numbers of fine-tuning epochs for CTW1500, Total-Text, MSRA-TD500 and ICDAR2017 were set to 200, 200, 250, and 400, respectively. 
Data augmentation techniques including random cropping, resizing, color variations, adding random noise, flipping and rotation were adopted. 

During the testing, following the existing practice, the input images were resized to $640\times 640$, $800\times 800$, $1024\times 1024$ and $1024\times 1024$ for CTW1500, Total-Text, MSRA-TD500 and ICDAR2017, respectively. 
The training was conducted on a single NVIDIA RTX 3090 GPU and the testing was conducted on a single NVIDIA Quadro P6000 GPU with a 3.60GHz Intel Xeon Gold 5122 CPU. 
Our proposed approach runs at a decent speed, with an average speed of 2.0 fps, 1.8 fps, 1.2 fps and 1.1 fps on the above tested datasets, respectively. 
Moreover, the processing speed is mainly limited by the non-maximum suppression process, which contributes over 60\% of the total processing time irrespective of the input sizes (see the Supplementary for more details).

\begin{table*}[t]
  \caption{The impact of the kernel size in DMOP and DMCL on F-measure.}
  \label{tab:SE}
  \centering
  \setlength\tabcolsep{2.6pt}
  \begin{tabular}{c|cccc|cccc}
    \hline
       Kernel Size & \quad & \quad CTW1500& \quad\quad & \quad& \quad & Total-Text \\  

	 \hline
		 DMOP, DMCL& Min F (\%) & Max F (\%) & Mean F (\%) & Std dev F  & Min F (\%) & Max F (\%) & Mean F(\%) & Std dev F\\
    \hline
   $2\times2,2\times2 $ &82.4&83.1&82.6&0.24  &82.0&82.2&82.1&0.09    \\
    $2\times2,3\times3$  
    &85.6&86.0&85.9&0.16  
    &86.6&86.9&86.7&0.10\\%
    $2\times2,4\times4$  
    &79.5&80.2&79.7&0.26    
    &82.0&82.7&82.4&0.26\\%
    $2\times2,5\times5$  
    &75.6&75.8&75.7&0.08    
    &75.6&76.0&75.7&0.14\\%
 \hline
 \hline

   $3\times3,2\times2$ &82.0&82.5&82.3&0.17
   &82.0&82.3&82.1&0.12\\%
   $3\times3,3\times3$ &85.7&85.9&85.8&0.09
   &86.0&86.4&86.2&0.16\\%
   $3\times3,4\times4$  &79.5&79.8&79.6&0.12    
   &81.5&81.8&81.6&0.11\\%
   $3\times3,5\times5$ &75.2&75.4&75.3&0.08    
   &75.3&75.6&75.5&0.10\\%
\hline
\hline
   $4\times4,2\times2$  &81.8&82.2&82.0&0.16
   &81.2&81.8&81.5&0.20\\%
   $4\times4,3\times3$  &83.1&83.5&83.3&0.15     
   &82.4&82.7&82.5&0.13\\%
   $4\times4,4\times4$  &79.0&79.5&79.3&0.17 
   &81.4&81.7&81.5&0.12\\%
   $4\times4,5\times5$  
   &75.0&75.4&75.3&0.15    
   &75.3&75.7&75.4&0.14\\%
\hline
\hline
   $5\times5,2\times2$ 
   &78.2&78.8&78.5&0.20  
   &77.2&77.8&77.6&0.21\\%
   $5\times5,3\times3$ 
   &82.1&82.4&82.3&0.11
   &82.0&82.4&82.2&0.14\\%
   $5\times5,4\times4$ 
   &77.1&77.7&77.4&0.19 
   &79.7&80.5&80.2&0.27\\%
   $5\times5,5\times5$ 
   &74.1&74.7&74.4&0.23   
   &74.7&75.0&74.9&0.11\\%
\hline

\end{tabular}
\end{table*}

\begin{table}[t]
  \caption{The impact of the residual connection on the detection.}
  \label{tab:res}
  \centering
  \setlength\tabcolsep{1.2pt}
  \begin{tabular}{c|ccc||ccc}
    \hline
       \multirow{2}{*}{Residual} & \multicolumn{3}{c}{CTW1500 } &  \multicolumn{3}{c}{Total-Text} \\  
		\cline{2-7}
		 &P (\%) & R (\%) & F (\%) & P (\%) & R (\%) & F (\%)\\
    \hline
    $\times$ &88.2&81.6 & 84.8   &87.2 &82.7& 84.9   \\
    $\checkmark$  &{\bfseries89.0[+0.8]}&{\bfseries83.2[+1.6]} & {\bfseries86.0[+1.2]}   &{\bfseries88.4[+1.2]} &{\bfseries85.5[+2.8]}& {\bfseries86.9[+2.0]}  \\%
  \hline
  
\end{tabular}
\end{table}

\subsection{Ablation Studies}
\label {subsect:ablationstudies}
\label{ablation}

\subsubsection {The Effectiveness of DMOP and DMCL}

To validate the effectiveness of our proposed DMOP and DMCL modules, we conducted ablation studies on the two most representative datasets of arbitrary-shape text detection, \textit{i.e.}, CTW1500 and Total-Text. Table~\ref{tab:1} shows the comparison results. 
In this table, `P', `R' and `F'  represent Precision, Recall, and F-measure, respectively, `DMOP' and `DMCL' denote our DMOP false detection removal branch and DMCL relational reasoning branch, respectively, 
and `OP' and `CL' denote the traditional morphological opening and closing, where the kernel size of OP and CL are $2\times2$ and $3\times3$, respectively, the same as those in our DMOP and DMCL. 
Here, the baseline is the model where our proposed DMCL is substituted with the traditional closing operation but without the DMOP noise suppression. Since our method follows a bottom-up design, a closing module, either CL or DMCL, is required to ensure the connectivity of the text segments.

From Table~\ref{tab:1}, the results show that our idea of using deep morphological operations to remove false detection and connect separated text segments is effective. Although the SEs and iteration selection of OP and CL are based on trial and error, the OP and CL adhere to the 
concept of our bottom-up design and can bring  $0.5\%$ and $0.5\%$ gains of F-measure on CTW1500 and Total-Text, respectively. 
When OP and CL are replaced by our proposed DMOP and DMCL, more significant improvements have been achieved on both datasets: $1.8\%$ and $1.3\%$, respectively. 
We attribute this performance gain to the guidance of the loss functions, which allows the DMOP and DMCL to be flexible and robust. To be specific, when CL is replaced by DMCL, the design idea of stretching text segments to fill the gap becomes trainable, and it brings $1.2\%$ and $0.8\%$ gains on CTW1500 and Total-Text, respectively.

The purpose of OP and DMOP is to remove false detections (spurious text-like objects), which clears the way for the subsequent relational reasoning. 
The effectiveness of this can be seen from the result of combining OP and DMCL, where even the traditional OP can remove some of the small text-like objects. The selected kernel of OP is a flat all-ones matrix with a size of $2\times2$. There is a performance gain of $1.3\%$ and $1.0\%$ on the two datasets. When OP is replaced by DMOP, the combination of DMOP and CL can also bring $0.8\%$ and $0.7\%$ gains on the datasets. 
The selected SE for CL here is a flat all-ones matrix with a size of $3\times3$. The DMOP contains trainable kernels and can decide whether to filter according to different situations of text instances. 
From the results shown in Table~\ref{tab:1}, we can conclude that our DMOP and DMCL modules are complementary. Through the gradient update process of MorphText, DMOP and DMCL complement each other to boost the final performance.

\subsubsection {The Impact of the SE Size}

In traditional morphological operations, the selection of SEs can greatly affect the final results. Here, we mainly test the impact of the size of SEs on the performance, since the values and patterns in the SE are learned with the neural network. 
In our experiments, the SEs of DMOP and DMCL were initialized with all-zero matrices. 
We chose the grid search strategy to search the optimal kernel size for DMOP and DMCL. The possible kernel sizes of DMOP were set to [$2\times2$, $3\times3$, $4\times4$, $5\times5$] and the possible kernel sizes of DMCL were set to [$2\times2$, $3\times3$, $4\times4$, $5\times5$]. 
We repeated each combination for five times and report descriptive statistics such as $max$, $min$, $mean$ and $std$ in Table~\ref{tab:SE}. 
The results show that the SE sizes $2\times2$ and $3\times3$ have achieved the highest average F-measure of $85.9\%$ and $86.7\%$ on CTW1500 and Total-Text. 

Moreover, if we keep the SE size of DMOP to be $2\times2$, the performance decreases as the SE size of DMCL increases from $3\times3$. 
Interpreting the results, we can see that the F-measure is sensitive to the kernel size: for DMCL, when the kernel size is greater than or equal to $4\times4$, it may cause very close text instances to stick together and hence results in a fluctuation of F-measure on CTW1500 and Total-Text. 
For kernel sizes less than $3\times3$, sometimes the DMCL fails to deal with large gaps and holes. For DMOP, when the kernel size is equal to or larger than $5\times5$, over-correction occurs, also resulting in a decrease in performance. 
When the kernel size is larger than or equal to  $5\times5$, it is more likely to use the DMOP as a feature extractor and have more significant impact on rectifying the former CNN features. 

The results in~\cite{derain, mondal2019morphological,intomo, prmo} show that when deep morphological operations are used as the classifier or feature extractor, the SE size is usually greater than or equal to $5\times5$. 
Instead, in our approach, DMCL and DMOP are used to complement CNNs, and they work well for removing false detections, filling hole areas and smoothing the results of CNNs. 
A large SE tends to over-correct the detection results and is ineffective when filtering some local false detections.

\subsubsection {Comparison with the GCN-based Method}

We also compare the proposed DMOP and DMCL with a GCN-based method~\cite{11}. 
Here, we removed the DMOP module and replaced the DMCL with the GCN implementation in~\cite{9} from our framework in Fig.~\ref{fig:overall_structure}. 
We mentioned earlier that GCNs only perform link prediction and have no ability to deal with errors accumulated from visual representation. 
The results in Table~\ref{tab:gcn} show that GCN-based methods can also be improved by the OP and DMOP operations. Both OP and DMOP can remove some noise patterns but the latter one is guided by the loss. 
As shown in this table, with the help of DMOP, the GCN-based method can achieve a performance comparable to state-of-the-art (SOTA) with an F-measure of $85.2\%$. 
Moreover, the link prediction step of GCN can be fully replaced by DMCL to ensure the connectivity of text segments.
Here, DMCL allows 
text segments to stretch along the most significant orientation of the text instance, which helps to alleviate the problem when text instances are broken in the middle due to the limitations of CNN features and their fixed geometric form.

\begin{table}[t]
  \caption{The effectiveness of our proposed DMOP and DMCL on a GCN based method.}
  \label{tab:gcn}
   \centering
    \setlength\tabcolsep{5pt}
  \begin{tabular}{c|c|c|c|c|c|c|c}
    \hline
    Datasets& OP& GCN & DMOP &DMCL &P (\%) &R (\%) &F (\%)\\
     \hline
    $\quad$&$ \times $&$\checkmark$&$\times$&$\times$&86.2 &83.2& 84.7\\
  
    $\quad$&$\checkmark$&$\checkmark$&$\times$&$\times$&86.9 &82.9& 84.9\\
  
    CTW1500&$\times$&$\checkmark$&$\checkmark$&$\times$&87.4 &83.2& 85.2\\
    $\quad$&$\times$&$\times$&$\checkmark$&$\checkmark$&{\bfseries89.0} &83.2 &{\bfseries86.0}\\
    \hline
\end{tabular}
\end{table}

\subsubsection {The Impact of the Residual Connection}
 
To verify the impact of the residual connection, we keep the kernel size and layers of DMOP and DMCL. Table~\ref{tab:res} shows that when residual connection was removed, all of precision, recall and F-measure were decreased. This supports the claim that residual connection can alleviate the problem of over-correction.

\subsubsection {The Impact of the Number of Layers}

To verify the impact of the number of erosion and dilation layers in DMOP and DMCL on the performance, we chose the grid search strategy to search the optimal number of layers of erosion/dilation for DMOP and DMCL. 
Since the erosion and dilation operations often exhibit symmetrical appearance, the number of erosion and dilation layers in DMOP was selected from $[2,3,4,5]$ and the number of erosion and dilation layers in DMCL was selected from $[2,3,4,5]$. 
We repeated five times for each combination and reported descriptive statistics such as $max$, $min$, $mean$ and $std$. 
As shown in Table~\ref{tab:layers}, with the increase of the erosion and dilation layers in DMOP, there is a decreasing trend in the F-measure of the proposed method on CTW1500. 
Too many erosion and dilation layers may grossly change the feature representation of CNNs, but in our work, we hope the DMOP module does not over rectify the results of CNNs. For DMCL, four erosion and dilation layers often achieve better F-measure. 
Fewer than four layers of erosion and dilation may cause the DMCL module fail to handle large gaps.

\begin{table}[t]
  \caption{The impact of the number of layers of erosion and dilation in DMOP and DMCL on CTW1500.}
  \label{tab:layers}
  \centering
  \setlength\tabcolsep{2.0pt}
  \begin{tabular}{c|cccc}
    \hline
       Layer $l$ & \quad & \quad CTW1500& \quad\quad \\  

	 \hline
		 DMOP, DMCL& Min F (\%) & Max F (\%) & Mean F (\%) & Std dev F  \\
    \hline
   $l=2,l=2$ &84.7&85.2&84.9&0.17\\%
    $l=2,l=3$  &85.2&85.5&85.3&0.12 \\%
    $l=2,l=4$  &85.6 &86.0 & 85.9 &0.16\\%
    $l=2,l=5$ &85.4&85.7&85.5&0.10\\%
 \hline
 \hline
   $l=3,l=2$ &84.1&84.9&84.6&0.30\\%
   $l=3,l=3$ &85.1&85.4&85.2&0.11\\%
   $l=3,l=4$ &85.3&85.5&85.4&0.06\\%
   $l=3,l=5$ &85.0&85.4&85.2&0.16\\%
\hline
\hline
   $l=4,l=2$  &84.0&84.4&84.2&0.16\\%
   $l=4,l=3$  &84.2&84.8&84.6&0.22\\%
   $l=4,l=4$  &84.7&85.0&84.8&0.11\\%
   $l=4,l=5$  &84.6&84.9&84.7&0.12\\%
\hline
\hline
   $l=5,l=2$ &83.6&84.1&83.8&0.18\\%
   $l=5,l=3$ &83.4&83.9&83.6&0.17\\%
   $l=5,l=4$ &83.8&84.3&83.9&0.17\\%
   $l=5,l=5$ &83.3&83.8&83.6&0.21\\%
\hline
\end{tabular}
\end{table}

\subsubsection{Discussion on Regularizing False Detections}
\begin{figure*}[t]
   \includegraphics[width=0.95\linewidth]{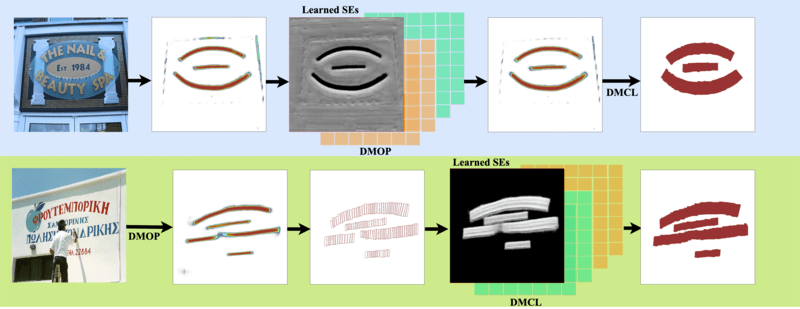} 
  \centering
  \caption{Visualisation of the intermediate results of MorphText addressing noise patterns (top) and the connection problem between text segments (bottom). }
  \label{fig:visualrobustness}
\end{figure*}

\begin{figure*}[t]
	\renewcommand{\tabcolsep}{1pt} 
	\renewcommand{\arraystretch}{0.8} 
	\centering
	\begin{tabular}{cccc}
	    \includegraphics[width=4cm,height=3cm]{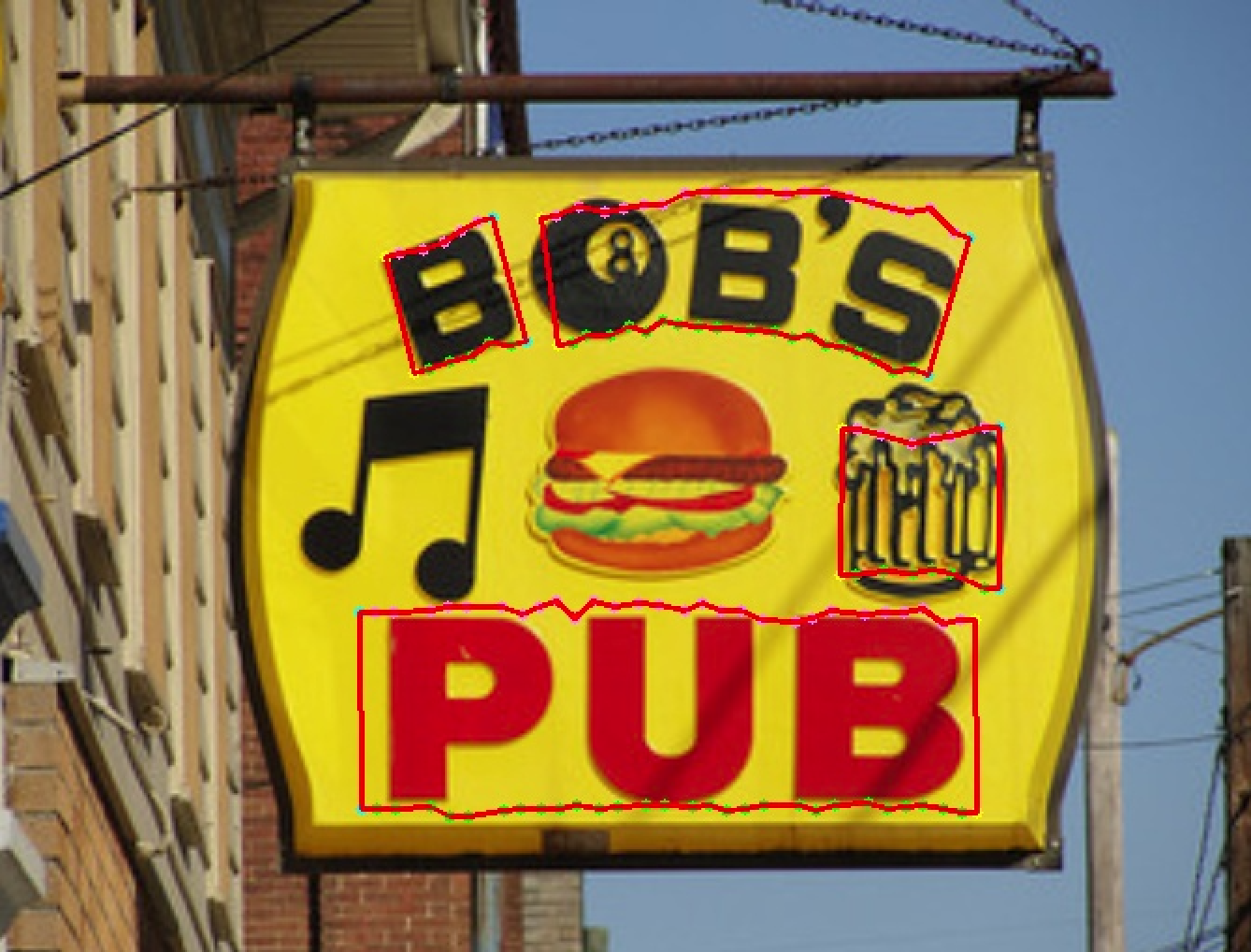} &
	    \includegraphics[width=4cm,height=3cm]{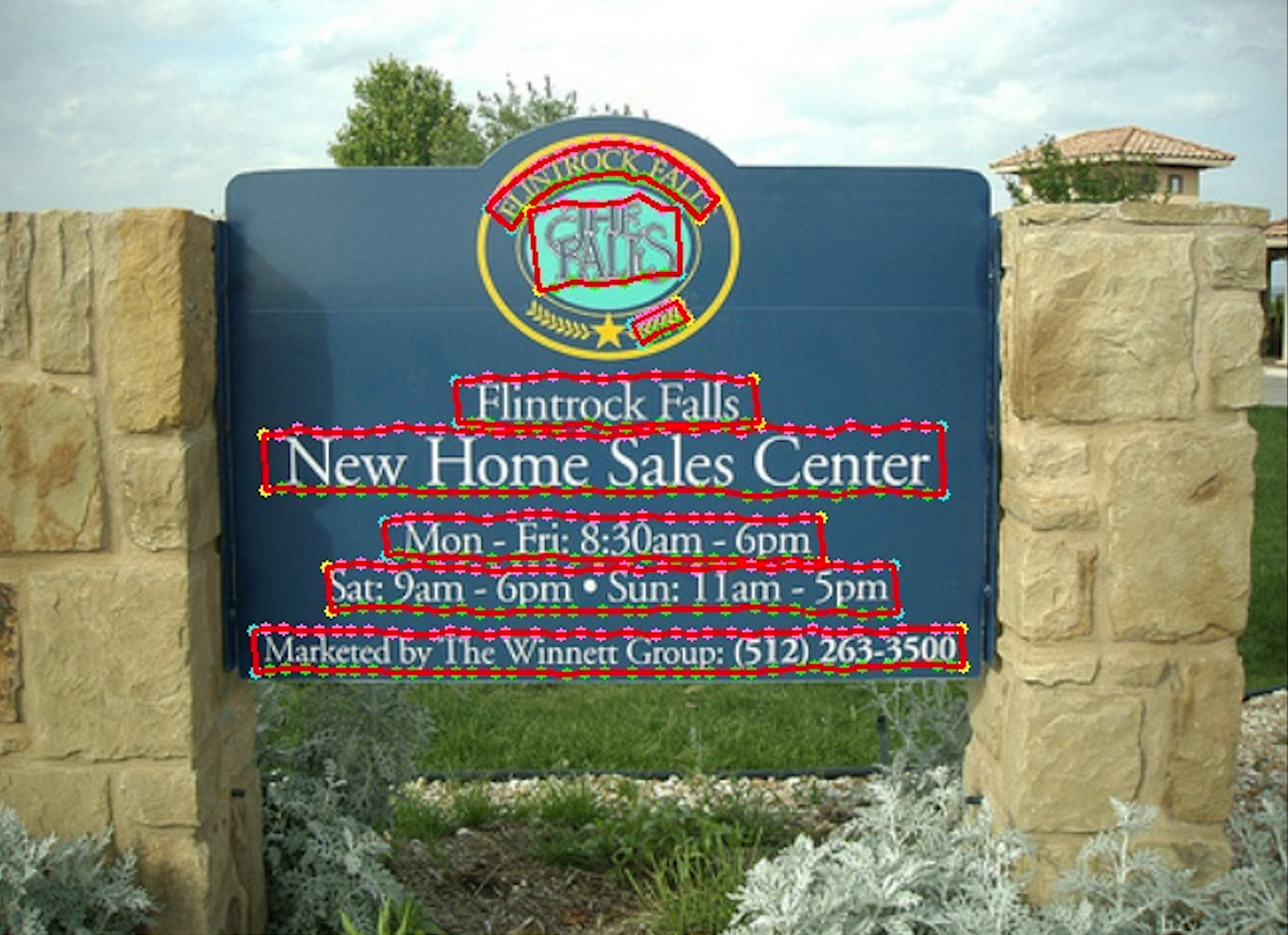} &
	    \includegraphics[width=4cm,height=3cm]{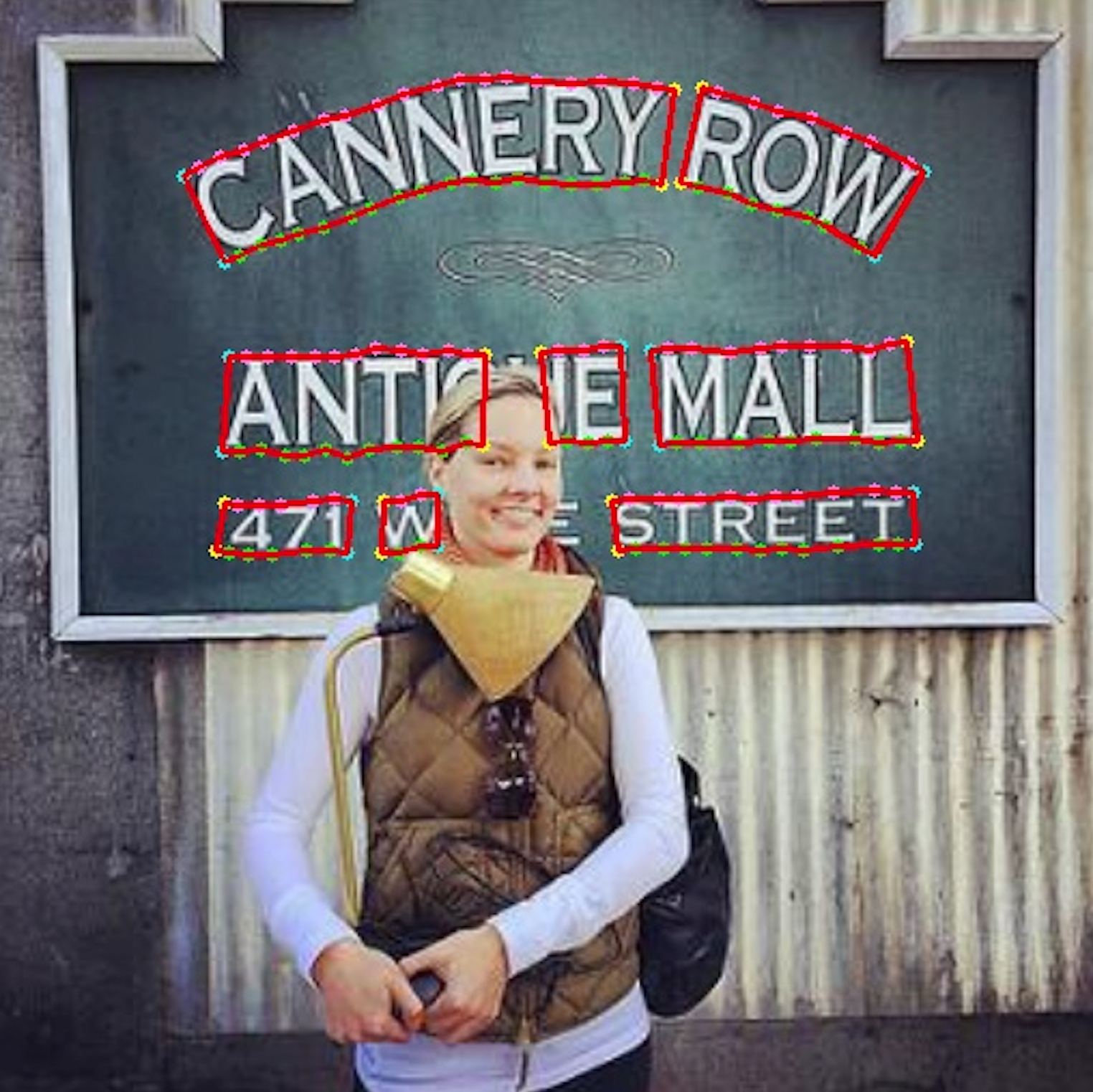} &
	    \includegraphics[width=4cm,height=3cm]{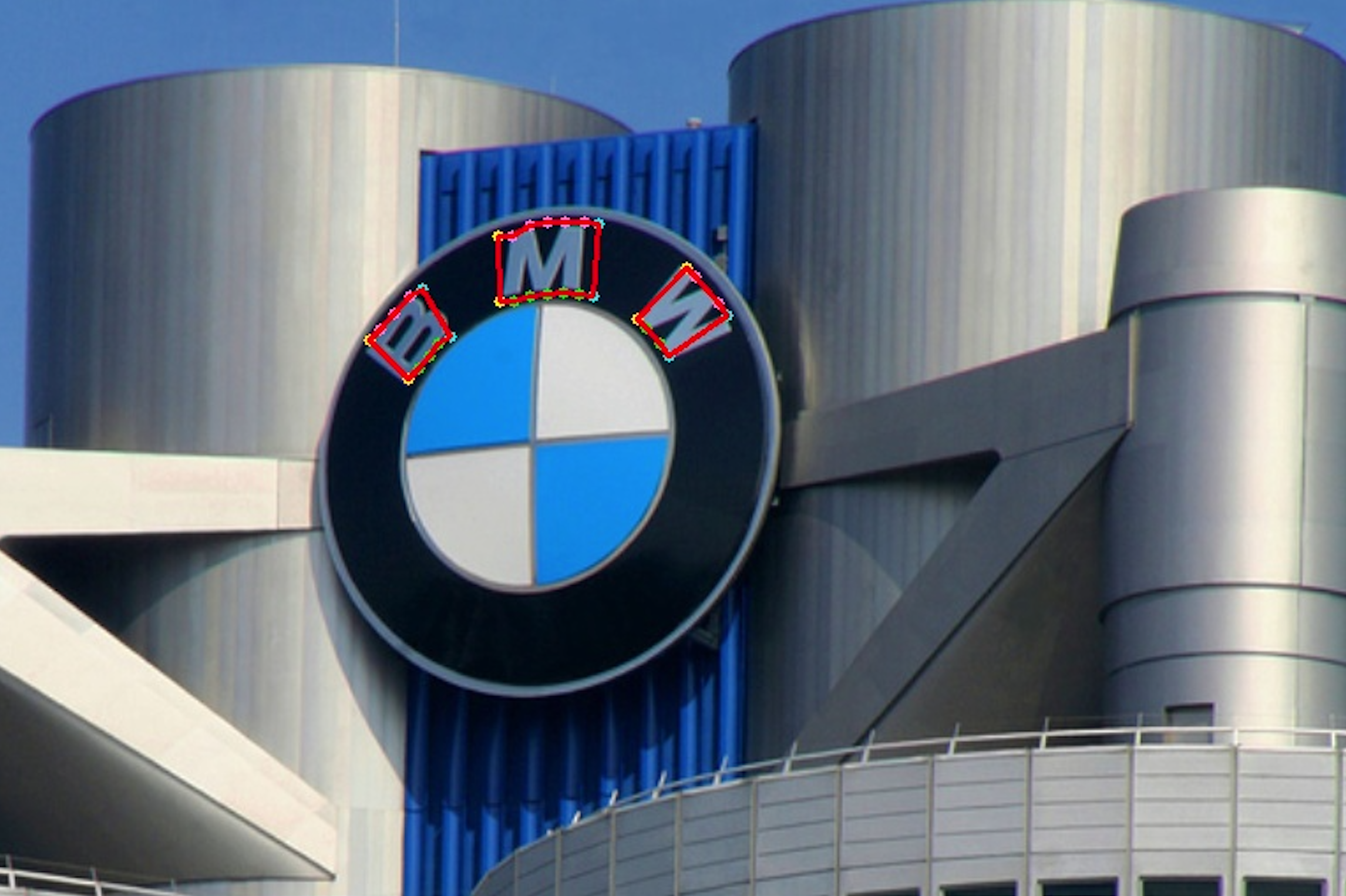}
		\\
		\includegraphics[width=4cm,height=3cm]{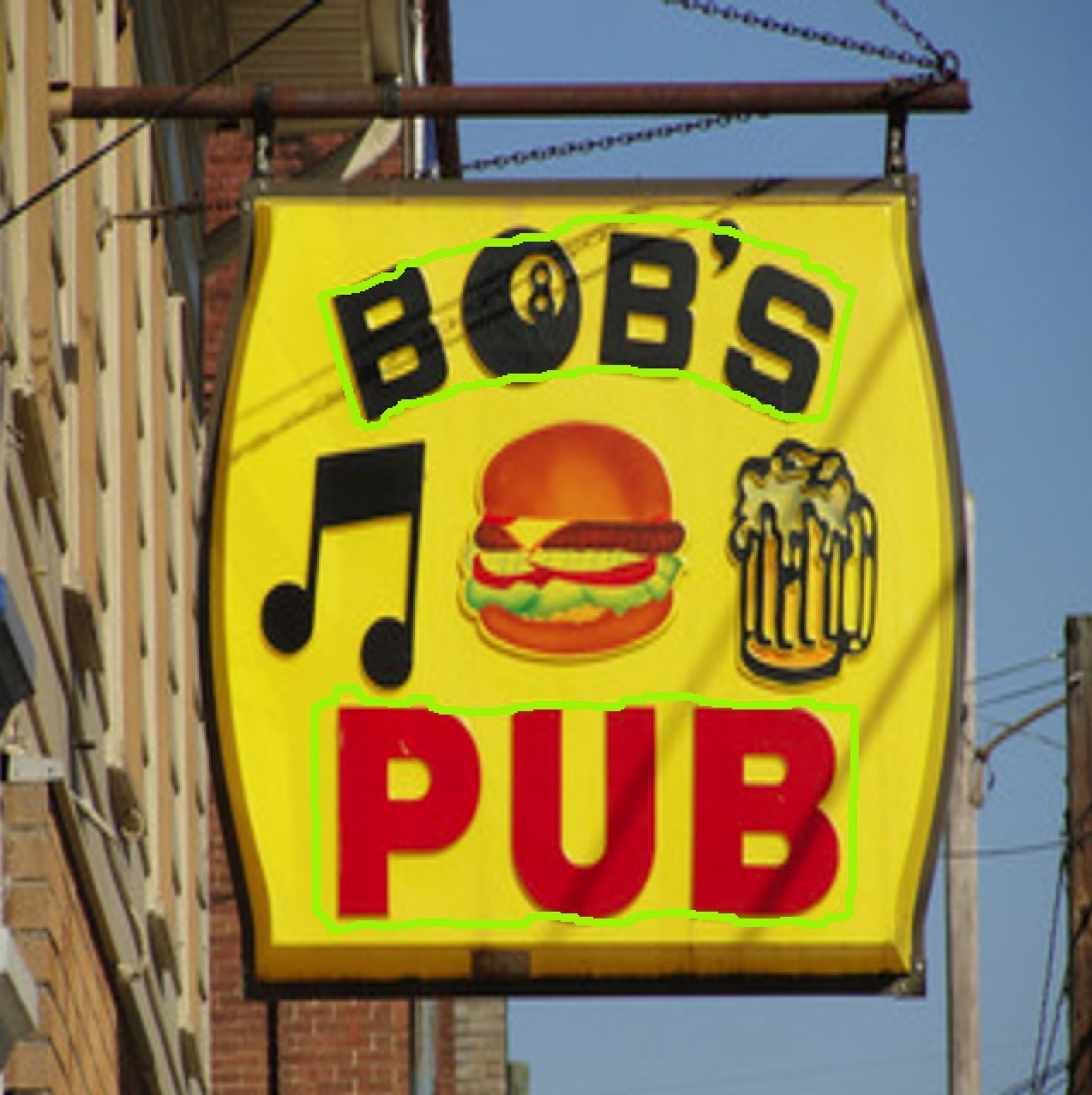} &
	    \includegraphics[width=4cm,height=3cm]{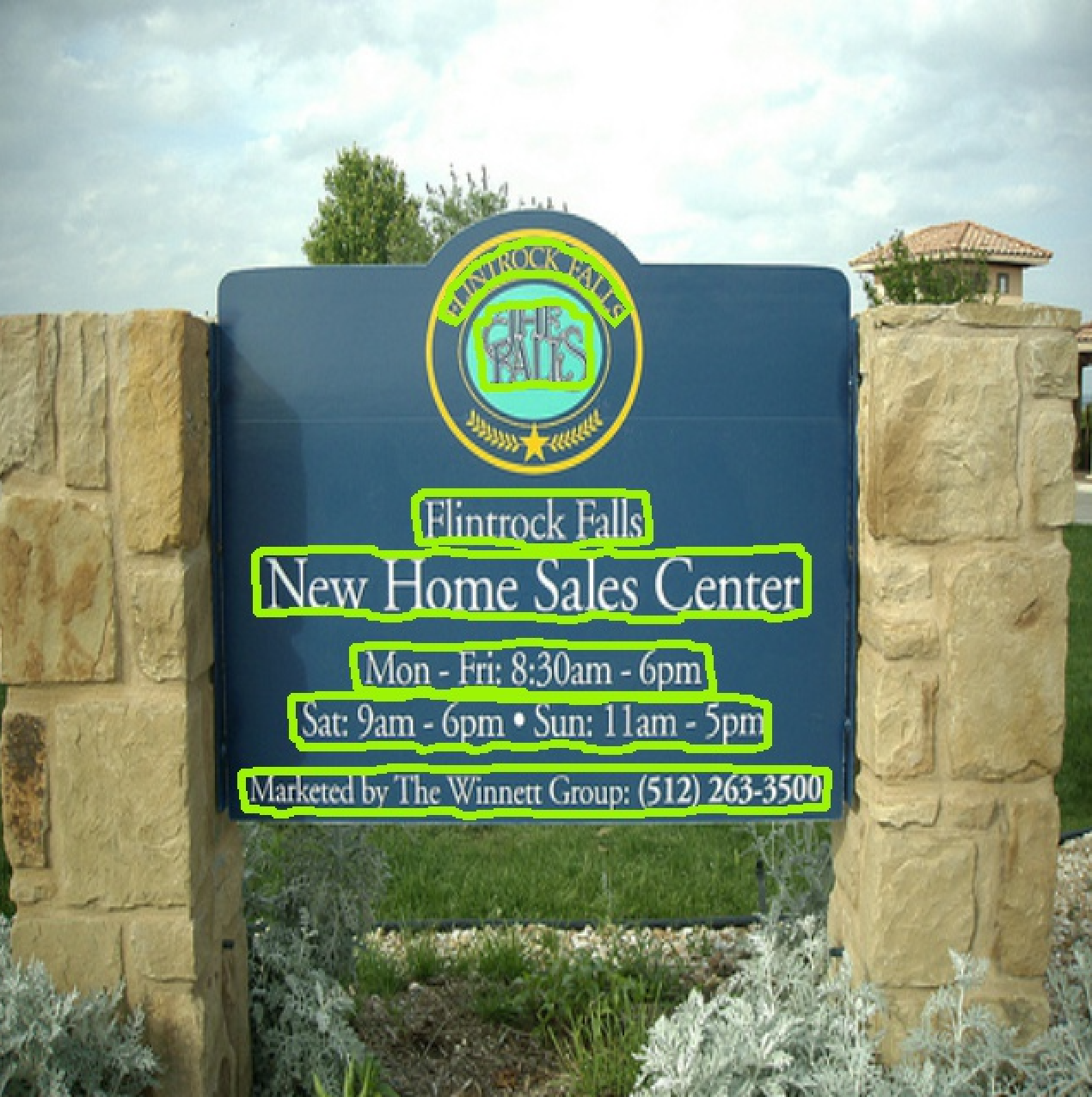} &
	    \includegraphics[width=4cm,height=3cm]{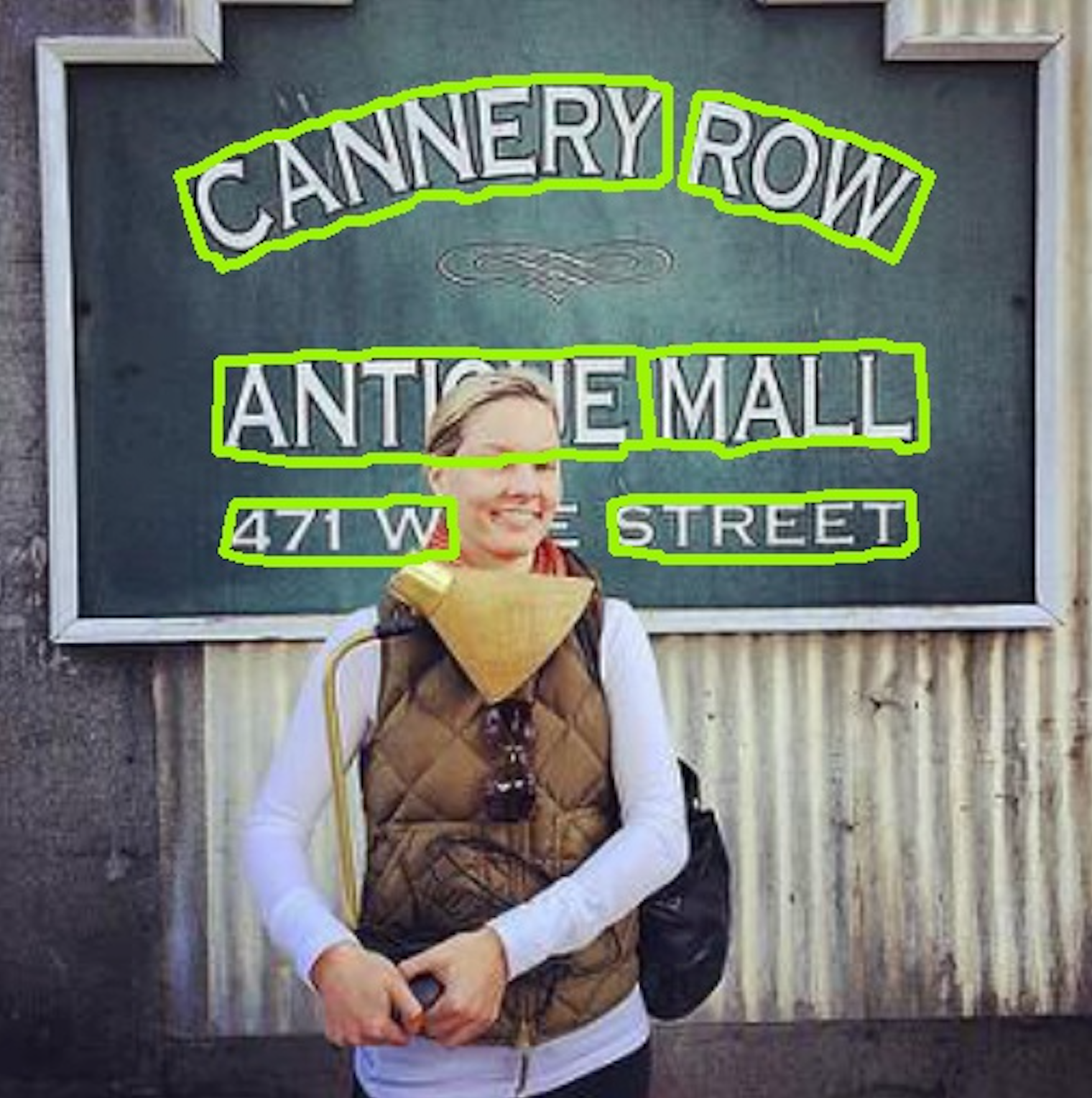} &
	    \includegraphics[width=4cm,height=3cm]{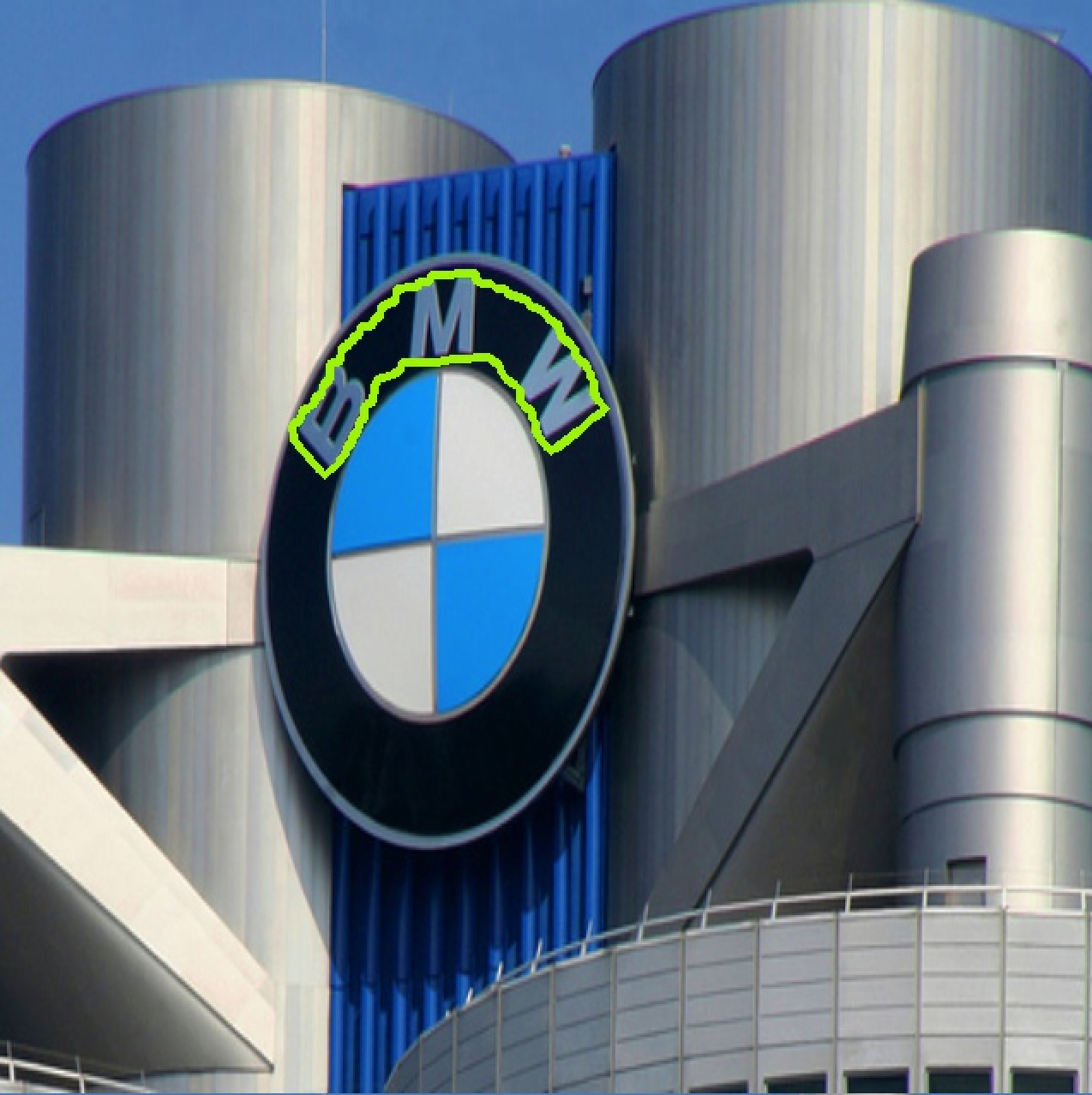}
		\\
	\end{tabular}
	\vspace{-1mm}
	\caption{Qualitative comparisons with the SOTA bottom-up method~\cite{11} (top row) on handling varied sizes of interfering patterns and relative large gaps}
	\label{fig:noise and link}
\end{figure*}

\begin{table}[t]
  \caption{The effectiveness and robustness of the proposed DMOP/DMCL on regularizing false positives and false negatives on CTW1500. }
  \label{tab:noise}
  \centering
  \begin{tabular}{c|ccc|ccc}
    \hline
  
		 DMOP/DMCL&TPs & FPs  & FNs  & P (\%) & R (\%) & F (\%)\\
    \hline
    $Without$ &2537& 423 & 531   &85.7 &82.7& 84.2   \\
    $2\times2,3\times3$  &{\bfseries2553$\uparrow$}&{\bfseries315$\downarrow$} & {\bfseries515$\downarrow$}   &{\bfseries89.0$\uparrow$} &83.2$\uparrow$& {\bfseries86.0$\uparrow$}  \\%
    
    $2\times2,4\times4$  &2429$\downarrow$&614$\uparrow$ & 639$\uparrow$   &79.8$\downarrow$ &79.2$\downarrow$& 79.5$\downarrow$  \\%
    
    $2\times2,5\times5$  &2313$\downarrow$&706$\uparrow$ & 755$\uparrow$& 76.6$\downarrow$ &75.4$\downarrow$& 75.6$\downarrow$  \\%
  \hline
  \hline
      $3\times3,3\times3$  &2571$\uparrow$&344$\downarrow$ & 497$\downarrow$  &88.2$\uparrow$ &{\bfseries83.8$\uparrow$}& 85.9$\uparrow$ \\%
    $4\times4,3\times3$  &2497$\downarrow$&420$\downarrow$ & 571$\uparrow$  &85.6$\downarrow$ &81.4$\downarrow$& 83.4$\downarrow$  \\%
    $5\times5,3\times3$  &2460$\downarrow$&444$\uparrow$& 608$\uparrow$   &84.7$\downarrow$ &80.2$\downarrow$& 82.4$\downarrow$  \\%
\hline 
\end{tabular}
\end{table}
\add{
We further validate the robustness of our method in addressing false detections.
In Figs.~\ref{fig:visualrobustness} and~\ref{fig:noise and link}, we show some qualitative comparison of the intermediate results, illustrating how MorphText addresses the varied sizes of noise and large linkage problems. The visualization of the results show that the proposed kernel sizes of DMOP and DMCL are robust to different types of noise and relatively large gaps between text segments.
Moreover, the various interfering text-like noise patterns, as well as the large holes or gaps between text segments that cause missing connections are the main causes of false detections. 
The results in Table~\ref{tab:noise} show the effectiveness and robustness of our proposed DMOP and DMCL with kernel sizes of $2\times2$ and $3\times3$ for dealing with false positives (FPs) and false negatives (FNs). 
Since both FPs and FNs have decreased, we conclude that the DMOP/DMCL can reduce the false positive detections and address the connection problem between text segments. 
Moreover, an increase in the kernel size of DMOP/DMCL may have a negative impact on dealing with FPs and FNs.}

\subsection{Comparison on Benchmark Datasets}
\begin{table*}[h]
 \setlength\tabcolsep{2pt}
  \centering
  \caption{Results on CTW1500, TOTAL-TEXT, MSRA-TD500 and ICDAR2017. (\S top-down methods, \dag bottom-up methods)}
  \label{tab:all}
  \begin{tabular}{c|c|c|cccc|cccc|cccc|cccc}

    \hline
      \multirow{2}{*}{Method}&\multirow{2}{*}{Venue} &  \multirow{2}{*}{Backbone} & \multicolumn{4}{c|}{CTW1500} &  \multicolumn{4}{c|}{Total-Text}  &  \multicolumn{4}{c|}{MSRA-TD500} &  \multicolumn{4}{c}{ICDAR2017}\\ 
		\cline{4-19}
		 &&&EXT &P(\%) & R(\%) & F1(\%) & EXT &P(\%) & R(\%) & F1(\%)&EXT & P(\%) & R(\%) & F1(\%)&EXT & P(\%) & R(\%) & F1(\%) \\
    \hline
    Wang et al.\S \cite{ATRR}&CVPR'19&VGG16&$\times$& 80.1&80.2&80.1 &$\times$& 80.9&76.2&78.5 &$\times$& 85.2&82.1 &83.6&-& -&-&-\\
    LOMO\S \cite{16}      &CVPR'19&ResNet50&\checkmark&85.7 &76.5 &80.8 &\checkmark&88.6 &75.7&81.6 &-&- &- &- &\checkmark&78.8 &60.6 &68.5\\
    DB\S \cite{39}         &AAAI'20&ResNet50&\checkmark& 86.9 &80.2 &83.4 &\checkmark&87.1 &82.5 &84.7 &\checkmark& {\bfseries91.5} &79.2 &84.9 &\checkmark& {\bfseries83.1} &67.9 &74.7 \\
    Dai et al.\S \cite{tmm1}    &TMM'21&ResNet50&$\times$&86.2&80.4&83.2 &$\times$&85.4&81.2&83.2 &-&-&-&- &$\times$&79.5&66.8&72.6\\
    MS-CAFA \S \cite{tmm2}       &TMM'21&ResNet50&$\times$&85.7&85.1&85.4 &\checkmark&84.6&78.6&81.5 &-&-&-&- &-&-&-&-\\
    ContourNet\S \cite{10}  &CVPR'20&ResNet50&$\times$ & 85.7 &84.0 &84.8  &$\times$ &  86.9& 83.9 &85.4 &- & -& - &- &- & -& - &-\\
    TextPerceptron\S \cite{textpreceptron}  &AAAI'20&ResNet50&\checkmark& 87.5 &81.9 &84.6 &\checkmark& 88.8 &81.8 &85.2 &-& -&-&-\\
    TextFuseNet\S\cite{9}    &IJCAI'20&ResNet50&\checkmark& 85.0 &{\bfseries85.8} &85.4 &\checkmark& 87.5 & 83.2 &85.3 &-& -& - &- &-& -&-&- \\
    PCR \S \cite{42}         &CVPR'21&DLA34~\cite{dla}&\checkmark& 87.2 &82.3&84.7 &\checkmark& 88.5 &82.0&85.2 &\checkmark& 90.8 &83.5 & {\bfseries87.0} &-& -&-&-\\
    FCENet\S \cite{43}         &CVPR'21&ResNet50&$\times$& 87.6 &83.4&85.5 &$\times$& 89.3 &82.5&85.8&-& -& - &- &-& -&-&-\\
    \hline
    \hline
    
    TextSnake\dag\cite{21} &ECCV'18&VGG16&\checkmark &67.9 &85.3 & 75.6 &\checkmark &82.7 &74.5& 78.4&\checkmark &83.2 &73.9& 78.3&-& -&-&-\\
    PSENet\dag\cite{psenet} &CVPR'19&ResNet50&\checkmark& 84.8 &79.7 &82.2 &\checkmark& 84.0 &78.0 &80.9 &-& - &- &- &\checkmark& 73.7 &68.2 &70.9\\
    TextRay\dag \cite{12}    &MM'20&ResNet50&$\times$&82.8 &80.4 &81.6 &$\times$&83.5 &77.8 &80.6 &-&- &- &- &-&- &- &-\\
    OPOM\dag \cite{tmm3}         &TMM'20&ResNet50&\checkmark&85.1&80.8&82.9 &\checkmark&88.5&82.9&85.6 &\checkmark&86.0&83.4&84.7 &\checkmark&82.9&70.5&76.2\\
    CRAFT\dag \cite{6}      &CVPR'19&VGG16&\checkmark& 86.0 &81.1 &83.5 &\checkmark& 87.6& 79.9& 83.6 &\checkmark& 88.2& 78.2& 82.9 &\checkmark& 80.6& 68.2& 73.9\\
    TextDragon\dag \cite{29}  &ICCV'19&VGG16&\checkmark& 84.5 &82.8 &83.6 &\checkmark& 85.6 &75.7 &80.3 &-&  - &- &- &-&  - &- &-\\
    PuzzleNet\dag \cite{8}   &arXiv'20&ResNet50&\checkmark& 84.1 &84.7 &84.4 &\checkmark& - &-&- &\checkmark& 88.2 &83.5&85.8 &-&  - &- &-\\
    DRRG\dag \cite{11}         &CVPR'20&VGG16&\checkmark& 85.9 &83.0 &84.4 &\checkmark& 86.5& 84.9 &85.7 &\checkmark& 88.1& 82.3 &85.1 &\checkmark& 74.5& 61.0 &67.3\\
    ReLaText\dag \cite{7}    &PR'21&ResNet50&\checkmark& 86.2 &83.3 &84.8 &\checkmark& 84.8& 83.1 &84.0 &\checkmark& 90.5& 83.2 &86.7 &-& -& - &-\\

    \hline
     Ours\dag       & -&ResNet50&$\times$ &89.0 &83.2& 86.0 &$\times$ &88.4 &{\bfseries85.5}& 86.9 &$\times$ &88.5 &82.7&85.5 &$\times$ &81.9 &74.0&77.8\\
    *Ours\dag        &-&ResNet50&\checkmark& {\bfseries90.0} &83.3 &{\bfseries86.5} &\checkmark& {\bfseries90.6} &85.2 &{\bfseries87.8} &\checkmark& 90.7 &83.5 &{\bfseries87.0} &\checkmark& 82.8 &{\bfseries74.2} &{\bfseries78.3}\\
 \hline
\end{tabular}
\end{table*}

\begin{figure*}[t]
	\renewcommand{\tabcolsep}{1pt} 
	\renewcommand{\arraystretch}{0.8} 
	\centering
	\begin{tabular}{cccc}
	    \includegraphics[width=4cm,height=3cm]{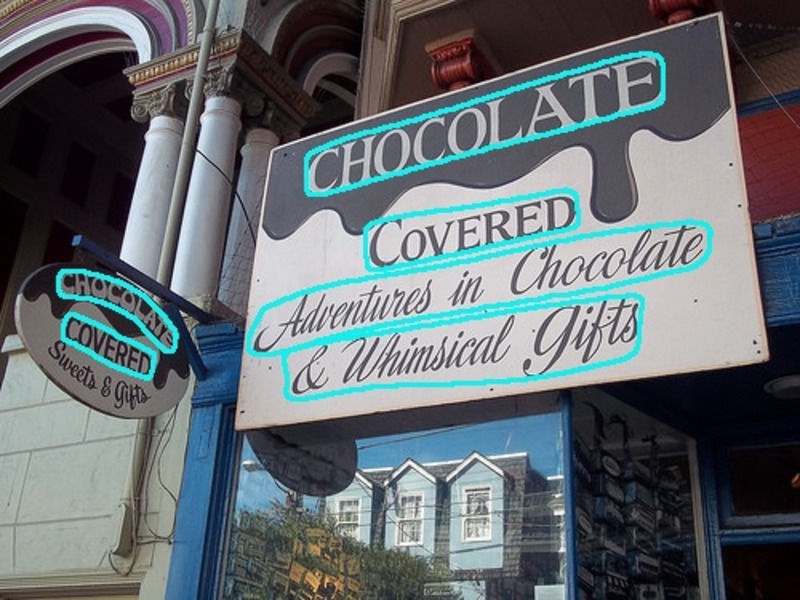} &
	    \includegraphics[width=4cm,height=3cm]{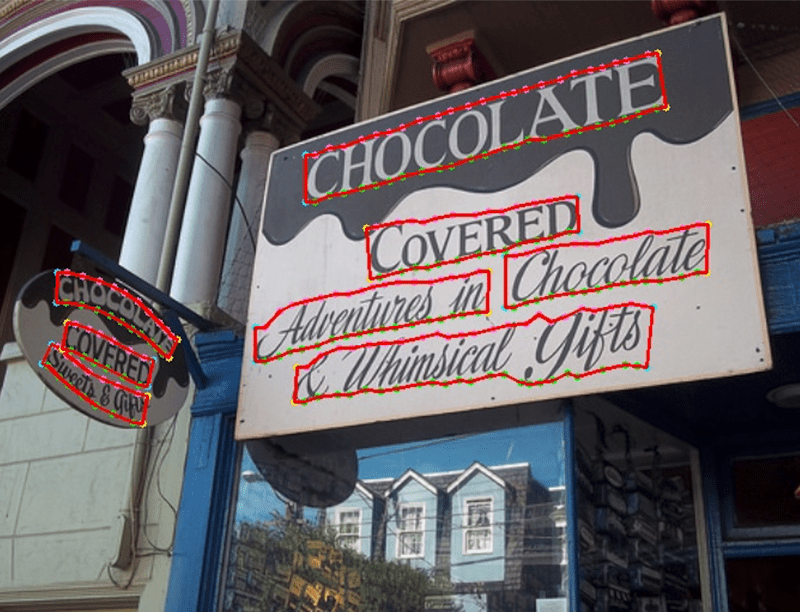} &
	    \includegraphics[width=4cm,height=3cm]{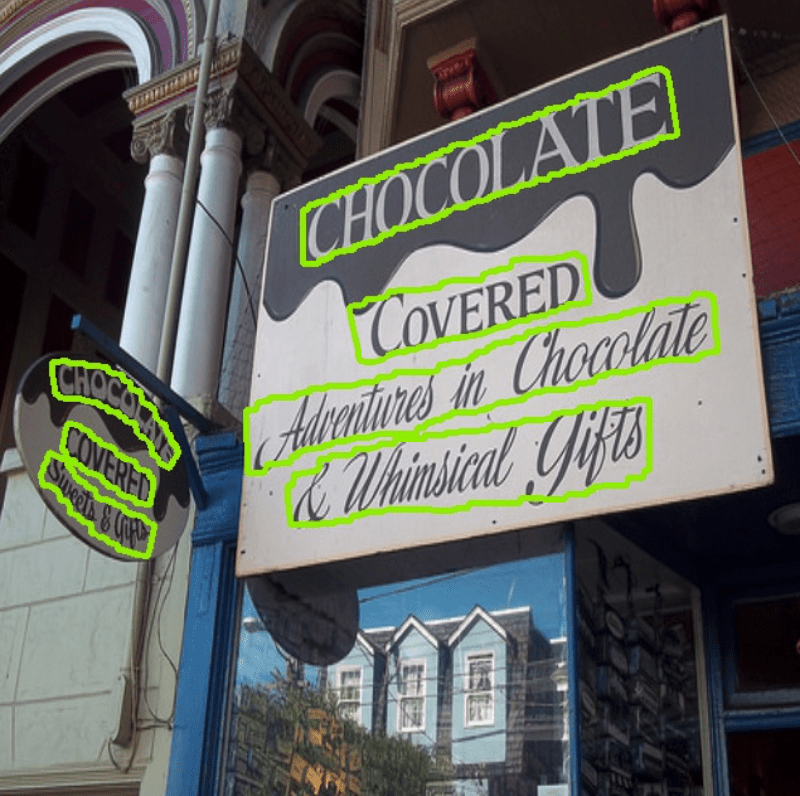} &
	    \includegraphics[width=4cm,height=3cm]{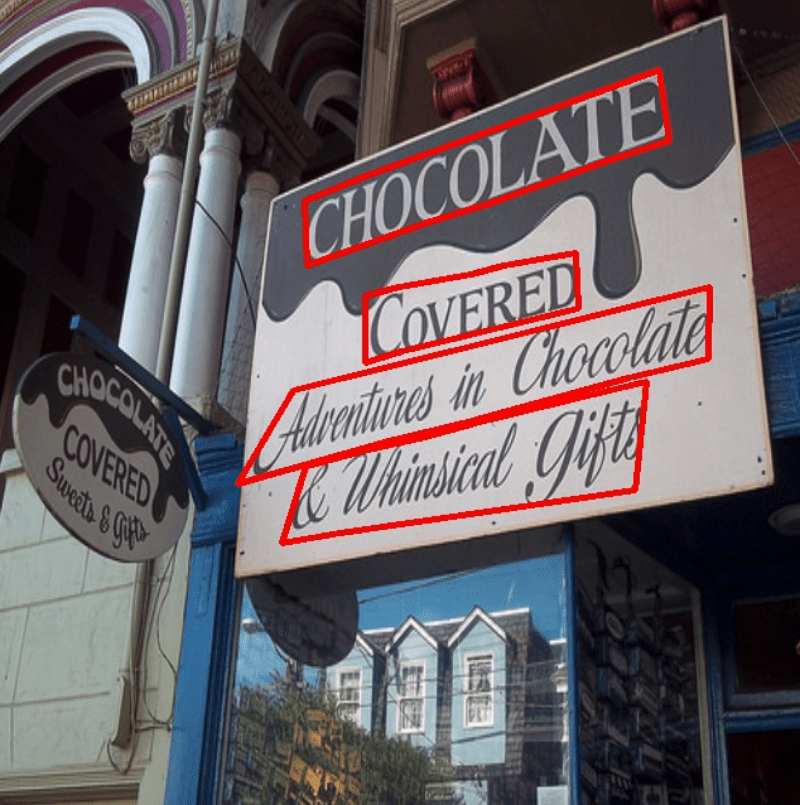}
		\\
		\includegraphics[width=4cm,height=3cm]{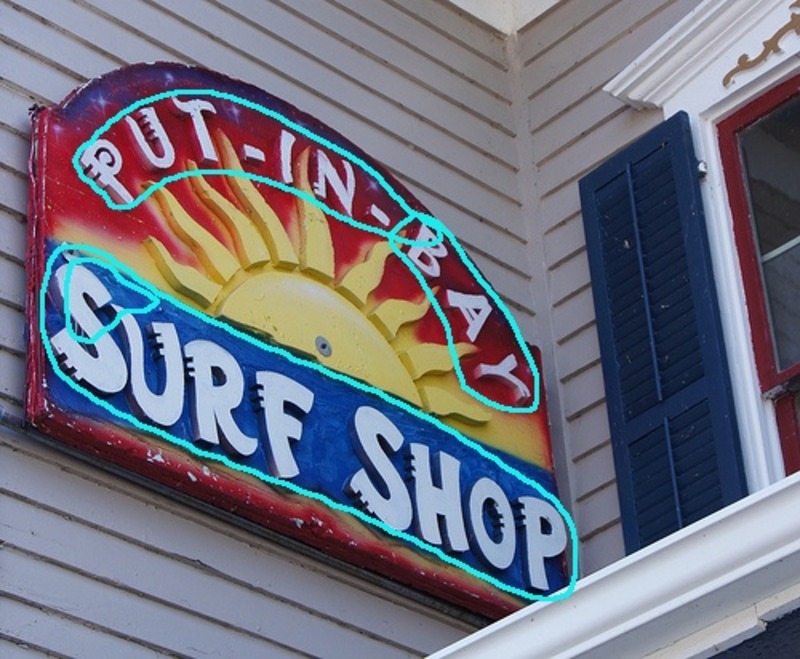} &
	    \includegraphics[width=4cm,height=3cm]{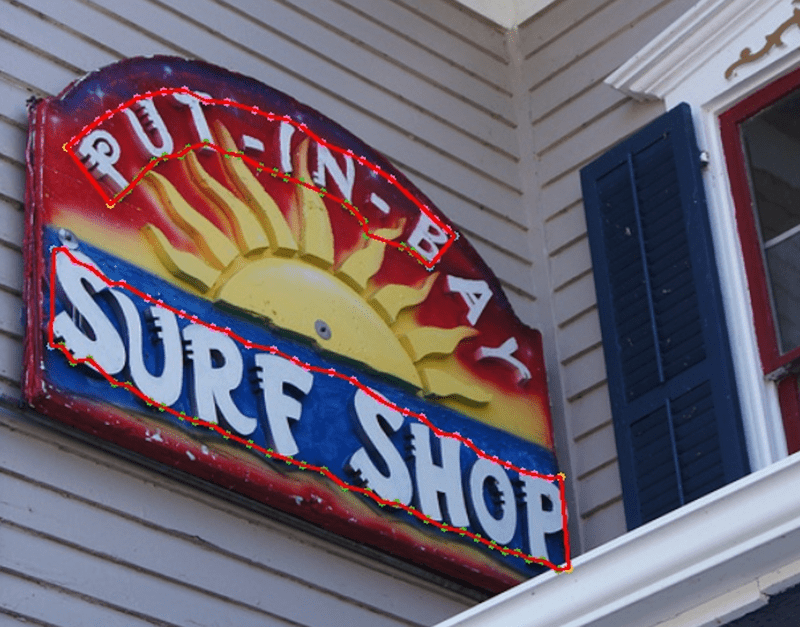} &
	    \includegraphics[width=4cm,height=3cm]{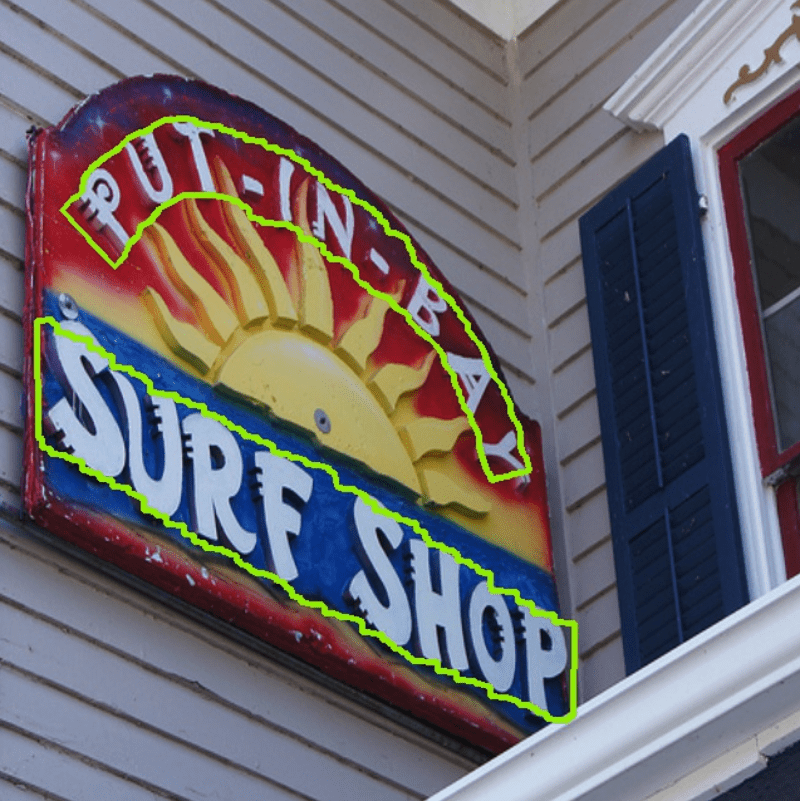} &
	    \includegraphics[width=4cm,height=3cm]{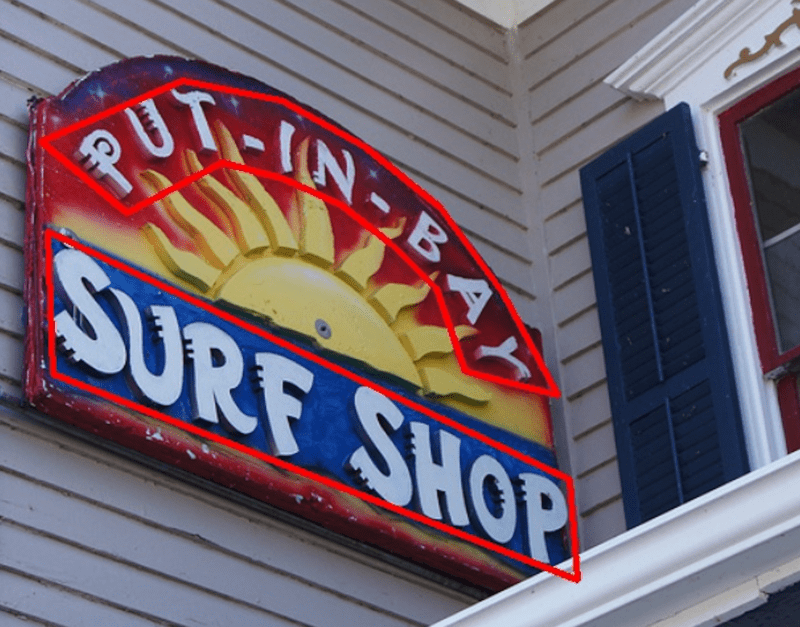}
		\\
		\includegraphics[width=4cm,height=3cm]{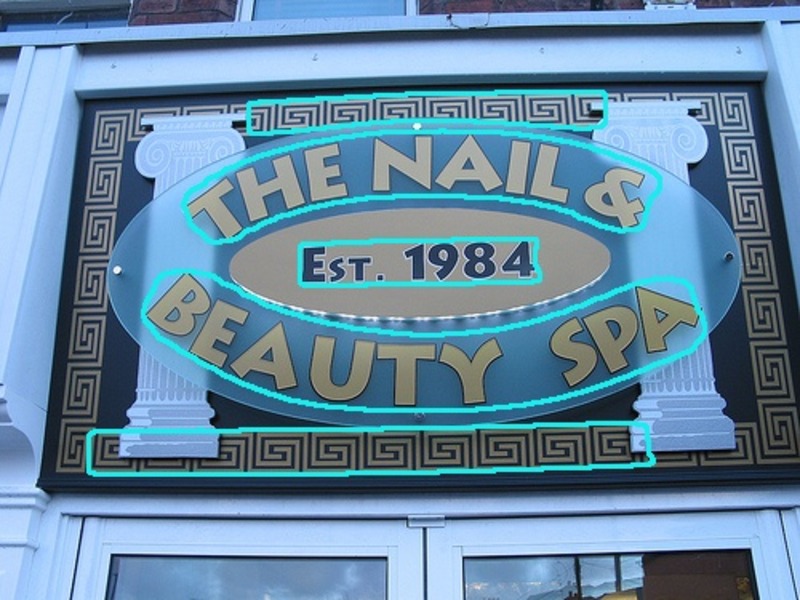} &
	    \includegraphics[width=4cm,height=3cm]{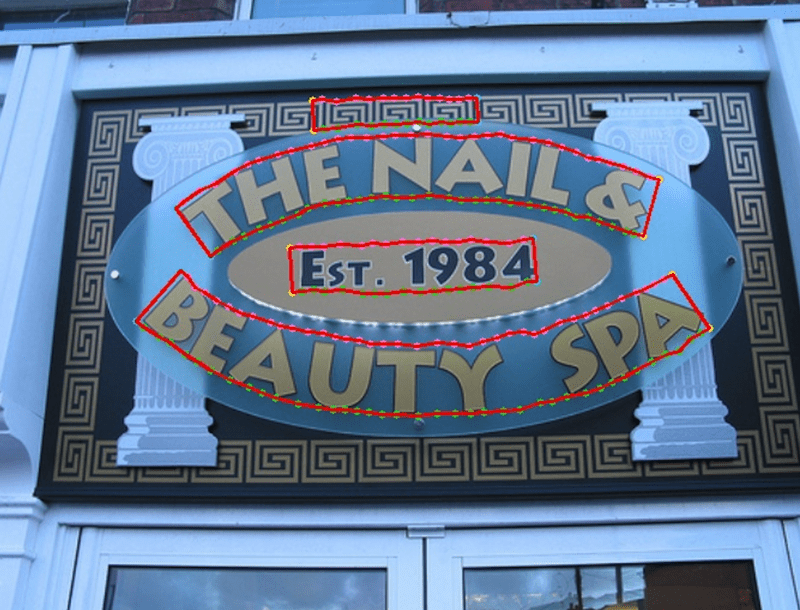} &
	    \includegraphics[width=4cm,height=3cm]{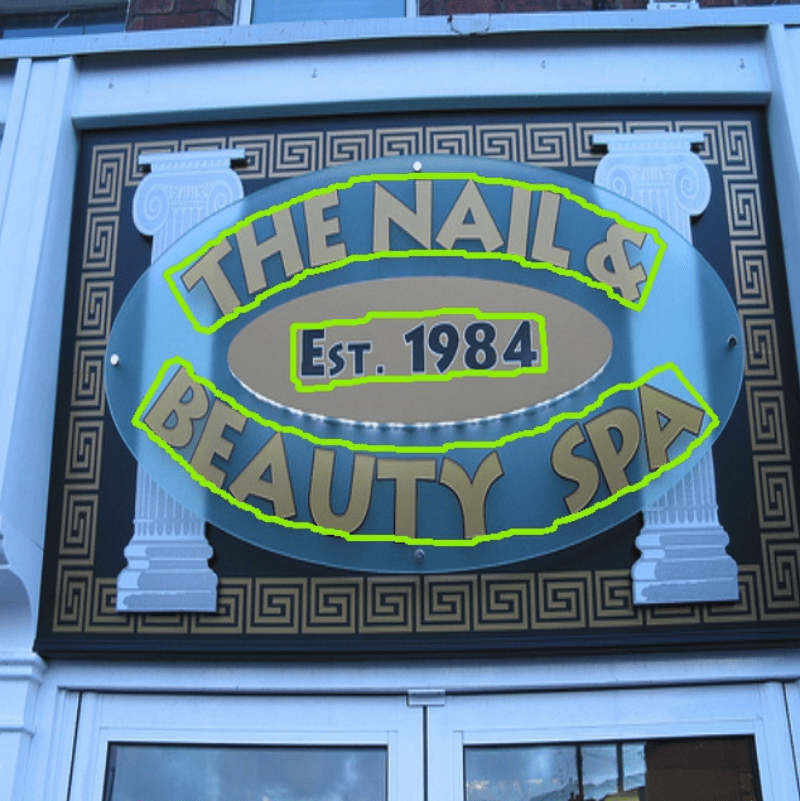} &
	    \includegraphics[width=4cm,height=3cm]{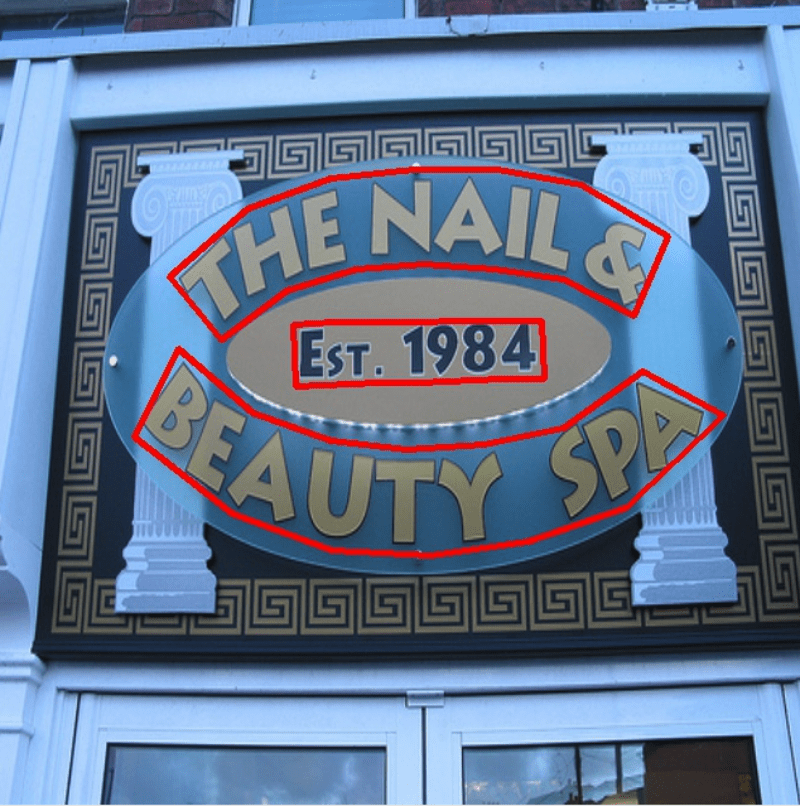}
		\\
		(a) TextFuseNet~\cite{9} & (b) DRRG~\cite{11} & (c) Ours & (d) Ground truth
		\\
	\end{tabular}
	\vspace{-1mm}
	\caption{Qualitative comparisons with the SOTA methods on challenging samples. }
	\label{fig:visualresult}
\end{figure*}

In this section, we present the experimental results conducted on the Total-Text, CTW1500, MSRA-TD500 and ICDAR2017 datasets. 
Some of the detection results of our proposed method are visualized in Fig.~\ref{fig:visualresults}.  
For fair comparison, networks with appropriate volumes such as VGG16 and ResNet50 were selected as the feature extraction backbones to exclude performance gains brought by the improvement of the backbones. 
Moreover, single-scale testing results are recorded to exclude performance gains brought by multi-scale testing. 

Table~\ref{tab:all} compares the detection results of MorphText with those of the SOTA approaches on the four benchmark datasets, respectively. 
The ``-" in the tables indicates that the comparative method has not reported their results on the dataset. The  ``EXT" in the tables indicates the comparative method has used external data for pre-training. 
\\

\noindent{\bfseries Comparison on arbitrary-shape datasets:} 
As shown in Table~\ref{tab:all}, we first compare our results with the SOTA methods on CTW1500. Our method has achieved {\bfseries90.0\%}, {\bfseries83.3\%} and {\bfseries86.5\%} in precision, recall and F-measure, and outperforms all of the comparison methods. 
Compared with the pioneering bottom-up methods~\cite{7,11}, our model has improved the detection accuracy by 1.7 percent. This result demonstrates that our DMCL module can replace the complex GCN reasoning process to make inferences on the relationships between text segments. 
When the DMOP module was applied to suppress false detections, the precision of our method significantly improved to 90.0\% on CTW1500. 
Our method remedies the error-prone post-processing steps in the bottom-up methods. 
The DMOP module introduces a further rectification under the guidance of the loss function, and it provides another way to remove false detections instead of introducing deformable or non-local convolutional blocks, as in the methods shown in~\cite{7,39,43,tmm1,tmm3}. 
Thanks to our proposed approach embedded with the deep morphological blocks, the results on CTW1500 demonstrate the superb effectiveness of MorphText when dealing with long curved texts.

As shown in Table~\ref{tab:all}, our method has achieved {\bfseries90.6\%}, {\bfseries85.2\%} and {\bfseries87.8\%} in precision, recall and F-measure on Total-Text, and outperforms the SOTA methods~\cite{43} by 2.0 percent in F-measure. 
The DMCL module can also handle short curved texts in Total-Text flexibly, breaking the text instance when it is appropriate to do so according to the loss.

Fig.~\ref{fig:visualresult} qualitatively compares the visual detection results of the proposed MorphText obtained on some samples containing long and curved text instances with results obtained using some recent top-down and bottom-up methods, namely TextFuseNet~\cite{9} and DRRG~\cite{11}. 
As shown in this comparison, 
our method exhibits greater robustness to interfering text-like noises patterns and 
is less prone to the connection problem
thanks to the proposed DMOP and DMCL modules.

Our results obtained on these two most representative arbitrary-shape text detection datasets demonstrate that the proposed deep morphological modules can handle the two important issues in bottom-up methods in an end-to-end manner, \textit{i.e.}, the unstable post-processing step and the missing connections. Our proposed DMOP module is able to remove the text-like objects from text segments, making DMCL's relational reasoning more reliable.
\\

\noindent{\bfseries  Comparison on multi-oriented datasets:} 
On the dataset MSRA-TD500, as shown in Table~\ref{tab:all}, our method has also achieved a comparable SOTA result of an F-measure of $87.0\%$, compared with the best-performing method PCR~\cite{42}. 
But different from~\cite{42}, which adopted the latest backbone~\cite{dla}, our MorphText only utilizes the original ResNet50 as the feature extraction backbone. 
Moreover, the modeling of our MorphText follows the regularity of texts, using the DMCL module to connect character-like text segments instead of introducing complex feature fusion process. 
Furthermore, different from the methods~\cite{8,7} that used multi-scale training, our method adopts a fixed scale to train the model. 
The results on MSRA-TD500 prove that the DMCL module can also link non-Latin text segments, allowing MorphText to deal with long multi-oriented texts in an end-to-end manner. 

The text detection results on the multi-lingual and multi-oriented dataset ICDAR2017 are shown in Table~\ref{tab:all}. 
As it shows, our MorphText has achieved {\bfseries82.8\%}, {\bfseries74.2\%} and {\bfseries78.3\%} in precision, recall and F-measure, so it also surpasses the best performing method~\cite{tmm3} by 2.1\% in F-measure. 
Text segments of non-Latin languages such as Chinese and Japanese are different to those of English, where non-Latin languages have no space between words. Therefore, connecting non-Latin text segments requires additional linkage information, since the intervals between non-Latin text segments vary considerably. 
Methods such as those in~\cite{6,9,7,11} attempted to introduce a character-level annotation as well as GCNs to guarantee the connectivity. 
Different from these methods, our method can directly regularize text segments by the proposed deep morphological modules, so the multi-lingual text segments are less likely to affect the performance. 
The highest F-measure and recall rate show the robustness of our approach in detecting multi-lingual texts.

\section{LIMITATION }

\begin{figure}[ht]
	\renewcommand{\tabcolsep}{1pt}
	\renewcommand{\arraystretch}{0.8} 
	\centering
	\begin{tabular}{ccc}
	    \includegraphics[width=2.7cm,height=2.2cm]{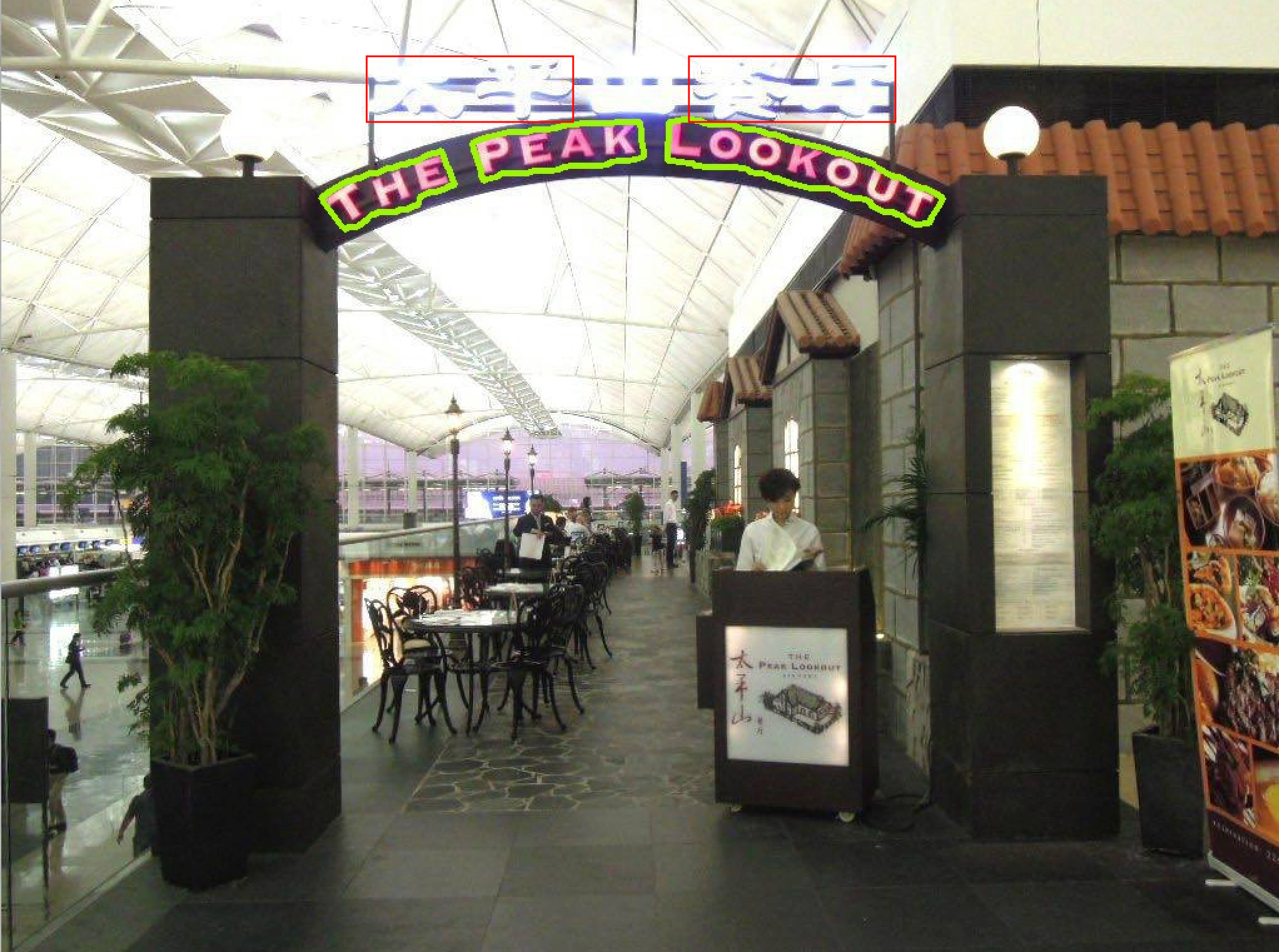} &
	    \includegraphics[width=2.7cm,height=2.2cm]{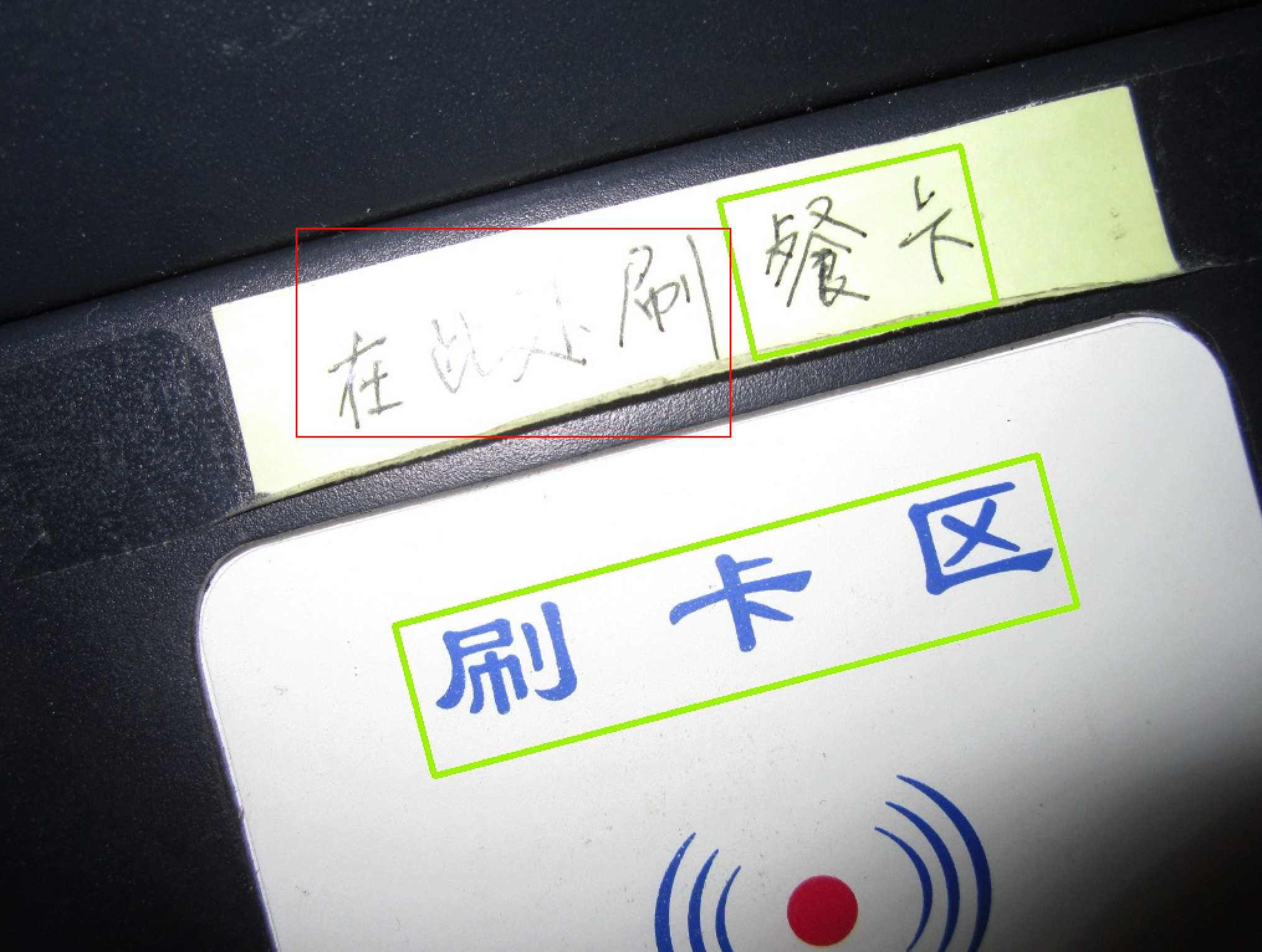} &
	    \includegraphics[width=2.7cm,height=2.2cm]{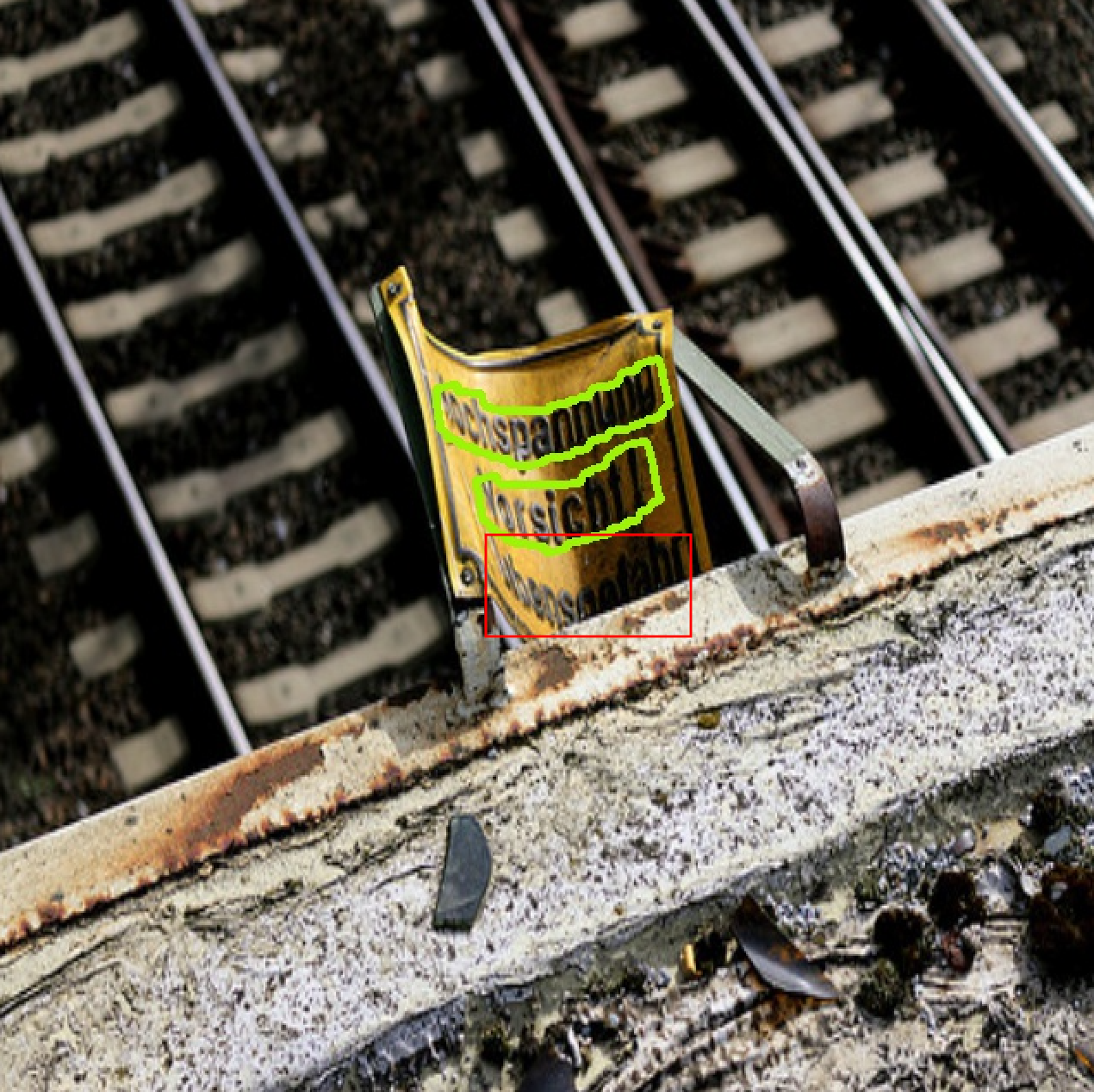} 
		\\ [1pt]
		\includegraphics[width=2.7cm,height=2.2cm]{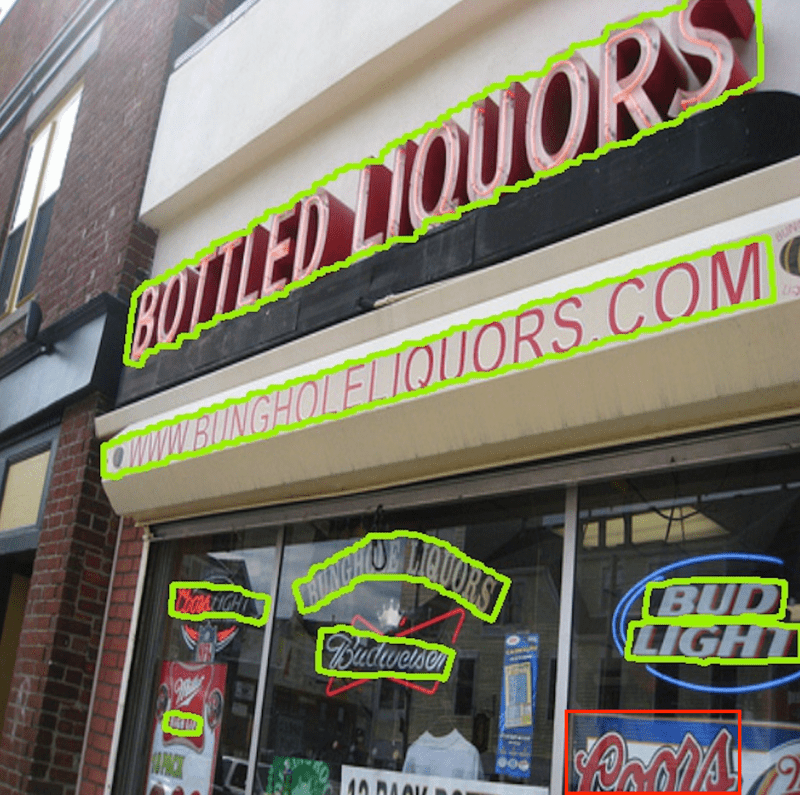} &
	    \includegraphics[width=2.7cm,height=2.2cm]{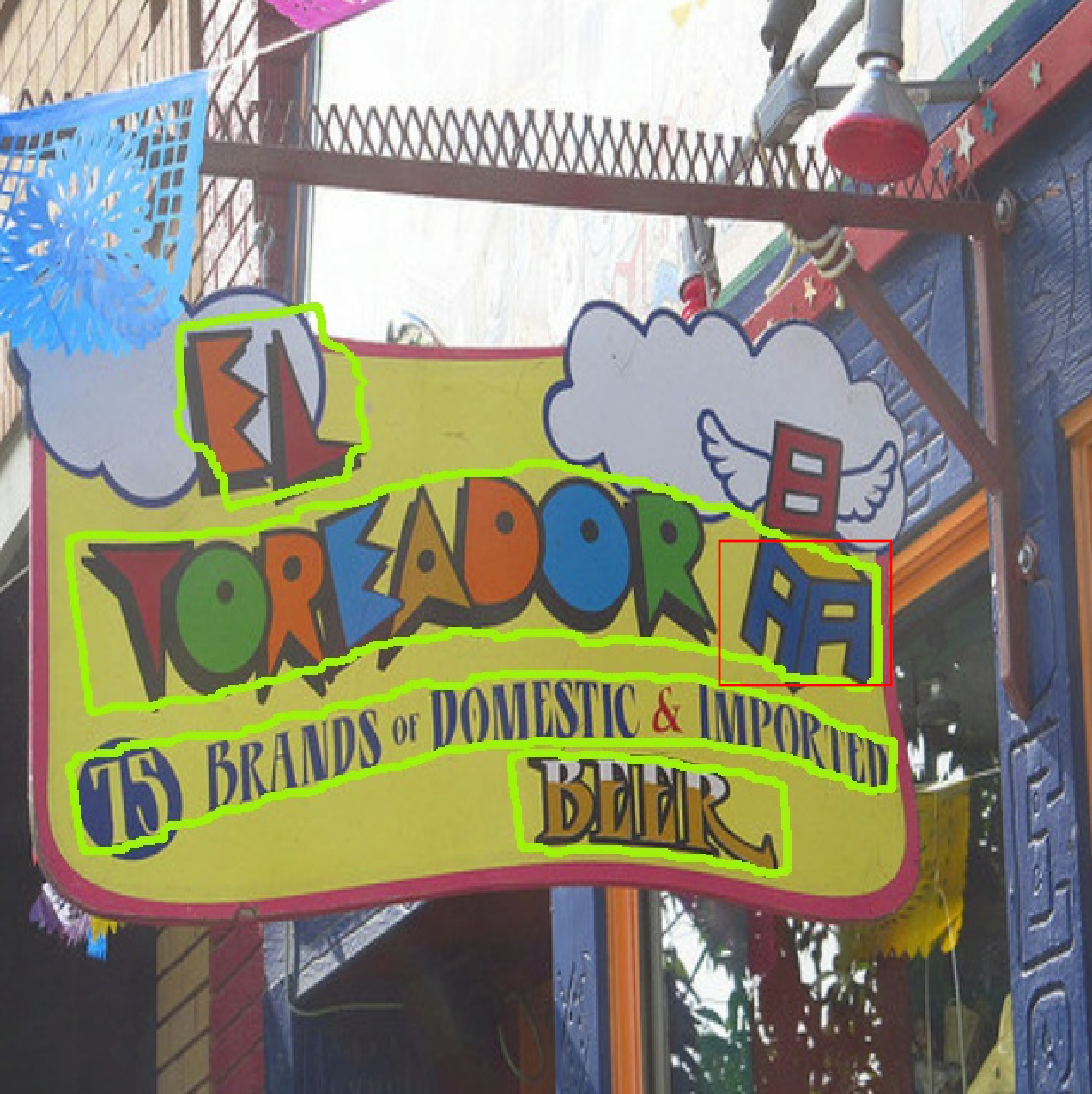} &
	    \includegraphics[width=2.7cm,height=2.2cm]{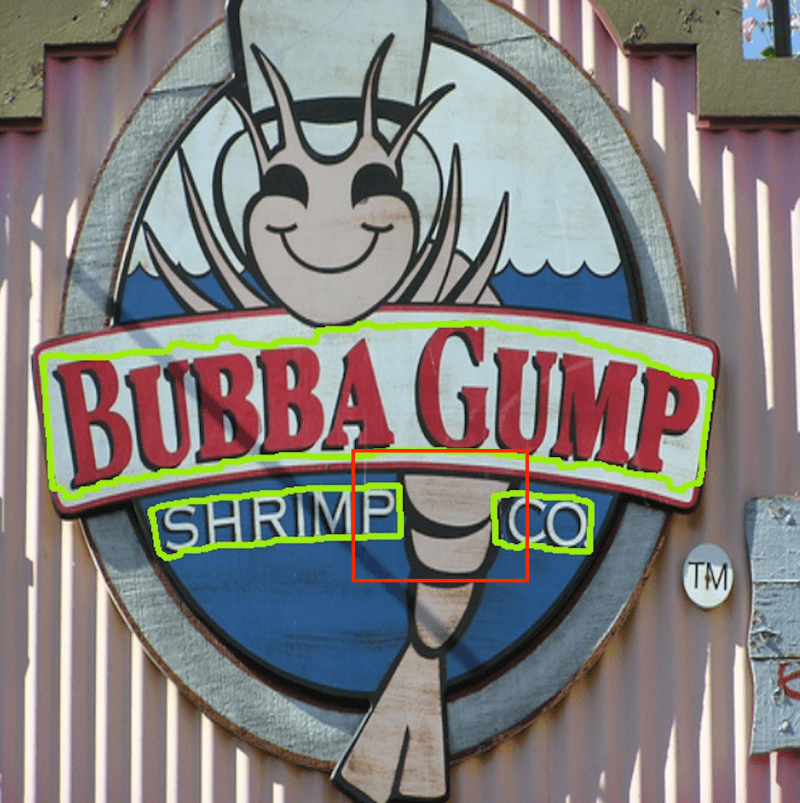} 
		\\ [1pt]
		\includegraphics[width=2.7cm,height=2.2cm]{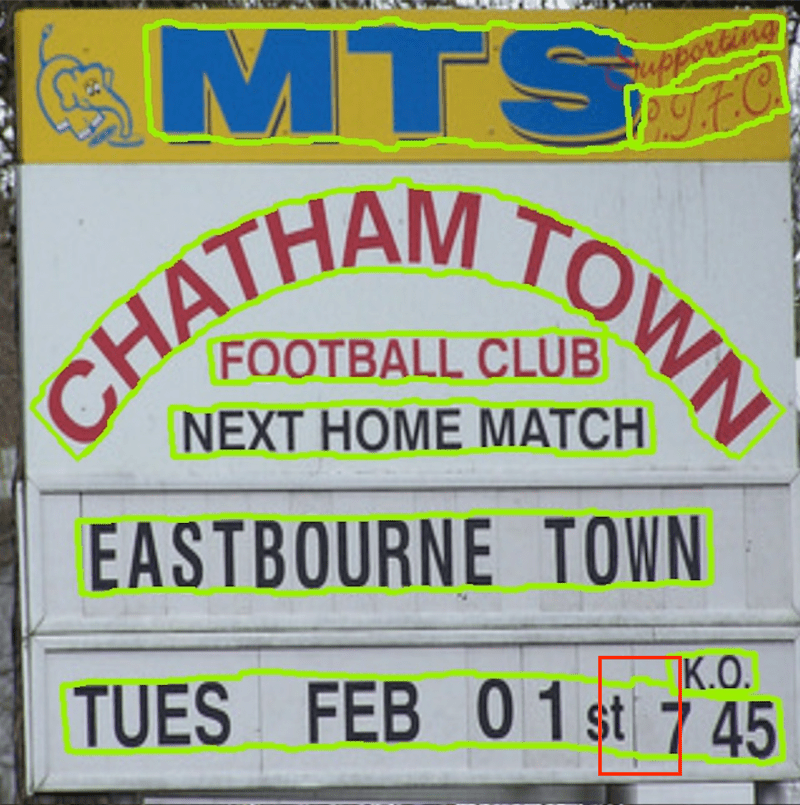} &
	    \includegraphics[width=2.7cm,height=2.2cm]{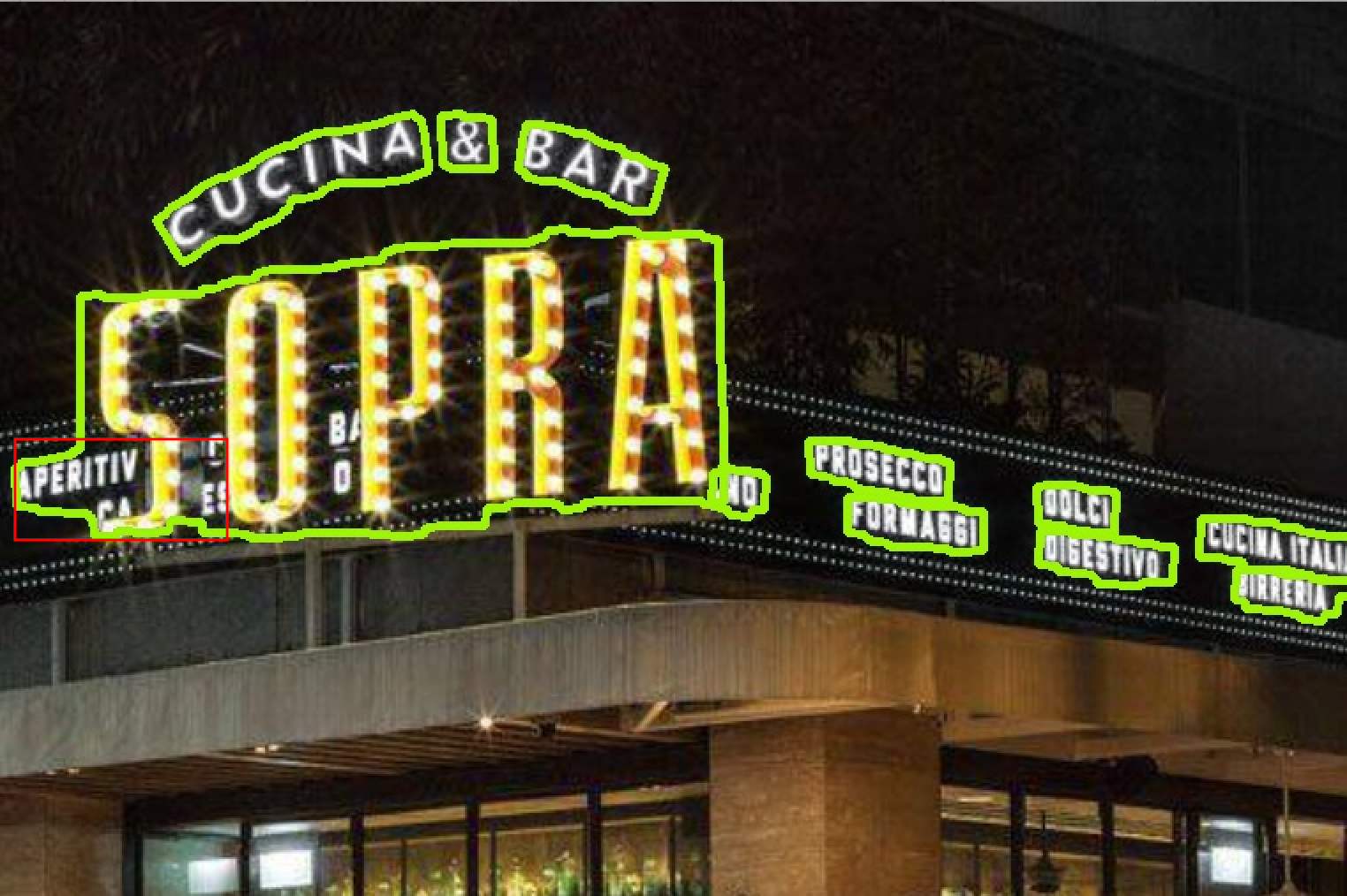} &
	    \includegraphics[width=2.7cm,height=2.2cm]{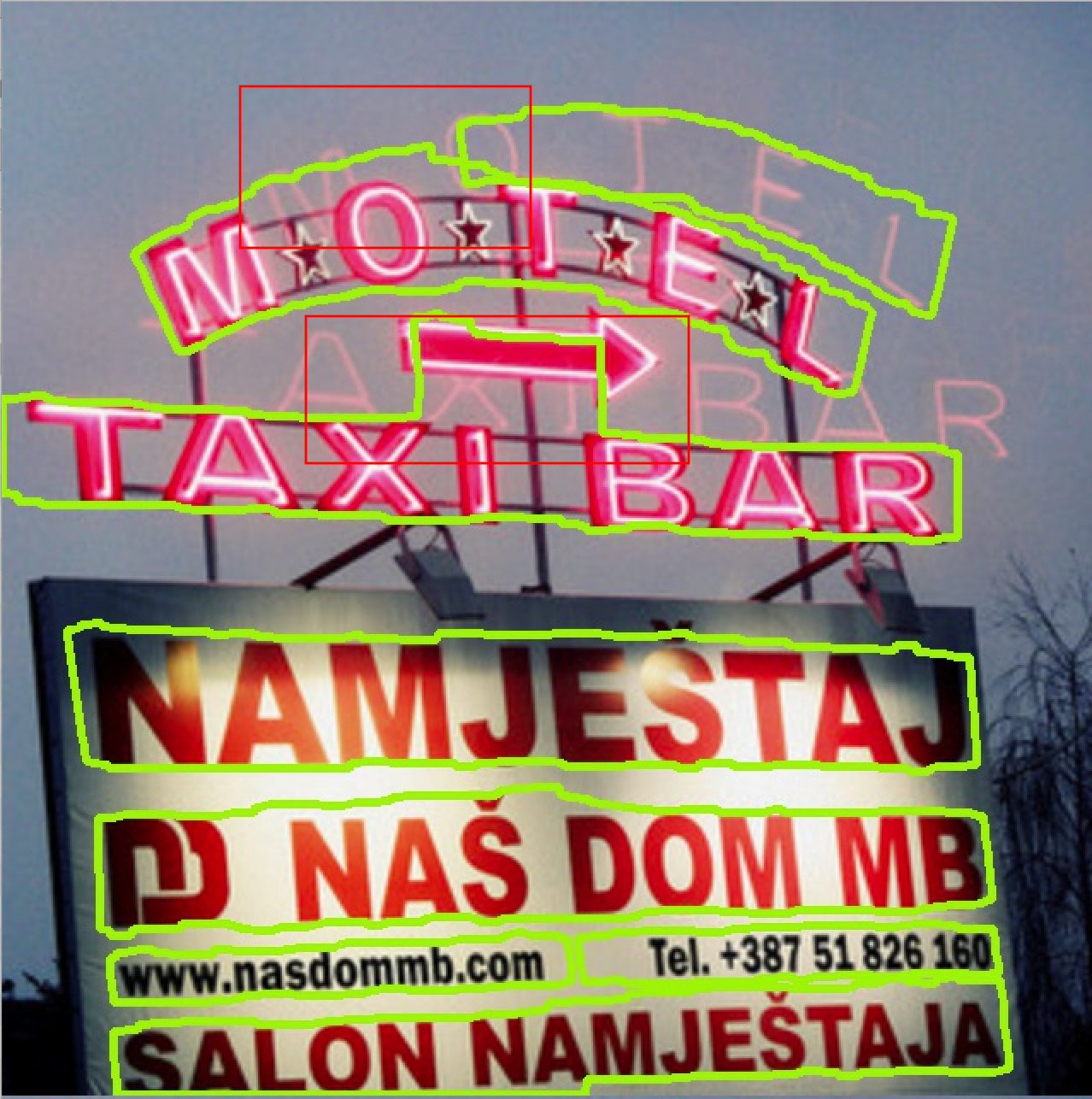} 
		\\
	\end{tabular}
	\vspace{-1mm}
	\caption{Some failure cases of the proposed MorphText, where the \add{green} bounding boxes indicate the detected results and the red bounding boxes highlight the failure areas.}
\label{fig:fail}
\end{figure}

Our method can alleviate the error accumulation problem and handle texts with larger word spaces, thanks to the newly proposed DMOP and DMCL modules. Some failure cases happen with low-contrast texts, object-like texts or extremely tiny texts such as the failure cases in the first five pictures in Fig.~\ref{fig:fail}, which are also a challenge for the SOTA methods~\cite{9,42,43}. 

For some confusing text segments such as the sixth and seventh pictures in Fig.~\ref{fig:fail}, it is very hard to determine the connectivity of text segments based purely on visual features, which is also challenging for the GCN-based methods~\cite{7,8,11}. 
Similar to~\cite{tmm1,tmm2}, our methods may also fail when text instances have large text overlapping with other texts, \textit{e.g.}, in the eighth and ninth pictures. 
Since overlapped texts are very rare in the training datasets,  our network may not learn the features of overlapped text.

\section{Conclusion}

In this paper, we have introduced deep morphology into the field of arbitrary-shape text detection for the first time, as an effective way to tackle the error accumulation of false text segment detection and the missing connection problems that prevent bottom-up text detection approaches from achieving their great potential for handling arbitrary-shape text. 
Two deep morphological modules have been designed and embedded into the network to regularize the text segments based on their patterns of regularity learned through training.
Extensive experiments conducted on four widely-used benchmark datasets have demonstrated the effectiveness of our proposed approach. 
The resulting state-of-the-art performance shows that deep morphology can be used for the challenging task of arbitrary-shape text detection and that the proposed methods have demonstrated an effective approach for the task.

\bibliographystyle{IEEEtran}
\bibliography{sample-base}

\end{document}